\documentclass[11pt]{article}
\usepackage{amssymb,amsthm,amsmath,float}
\usepackage[total={6.65in,8.75in}, top=1.2in, left=0.9in, includefoot]{geometry}
\usepackage{graphicx}
\usepackage{url}

\newtheorem{teo}{Theorem}

\newcommand{\Eq}[1]{(\ref{eq:#1})}
\newcommand{\Th}[1]{Thm.~\ref{thm:#1}}
\newcommand{\Sec}[1]{\S \ref{sec:#1}}
\newcommand{\Fig}[1]{Fig.~\ref{fig:#1}}

\newcommand{\InsertFig}[4]
{\begin{figure}[h!t]
       \centerline{
         \includegraphics[width=#4]{./figures/#1}
       }
       \caption{{\footnotesize  #2}
       \label{fig:#3}}
\end{figure}}

\newcommand{\InsertFigTwo}[5] {
\begin{figure}[ht]
       \centerline{
         \includegraphics[height=#5]{./figures/#1}
         \hskip 0.2in
         \includegraphics[height=#5]{./figures/#2}
       }
       \caption{{\footnotesize  #3}
       \label{fig:#4}}
\end{figure}}

\newcommand{\bN}{{\mathbb{ N}}}

\newcommand{\bR}{{\mathbb{ R}}}

\newcommand{\bT}{{\mathbb{ T}}}
\newcommand{\bZ}{{\mathbb{ Z}}}

\newcommand{\cL}{{\cal L}}
\newcommand{\cO}{{\cal O}}
\newcommand{\cR}{{\cal R}}

\newcommand{\cT}{{\cal T}}

\newcommand{\eps}{\varepsilon}
\newcommand{\vphi}{\varphi}

\newcommand{\fro} {Froeschl\'e }
\newcommand{\vol} {{\mathcal V}}  
\newcommand{\smooth}[2][T]{\langle{#2}\rangle_{\lambda}^{#1}}
\newcommand{\Fix}[1]{\mbox{Fix}(#1)}

\newcommand{\sgn}{\mathop{\rm sgn}\nolimits}
\newcommand{\Tr}{\mathop{\rm tr}}

\newcommand{\beq}[1]{\begin{equation}\label{eq:#1}}
\newcommand{\eeq}{\end{equation}}

\newenvironment{se}[1]{\equation\label{eq:#1}\aligned}{\endaligned\endequation}
\newcommand{\bsplit}[1]{\begin{se}{#1}}
\newcommand{\esplit}{\end{se}}

\title{Diffusion and Drift in Volume-Preserving Maps}
\author{N. Guillery and J.D.~Meiss     \\
 \begin{tabular}{c}
	Department of Applied Mathematics\\
    University of Colorado \\
	Boulder, CO 80309-0526 \\
	James.Meiss@colorado.edu\\ 
\end{tabular}
}
\date{\today}
\begin{document}
\maketitle
\begin{abstract}
 
A nearly-integrable dynamical system has a natural formulation in terms of  actions, $y$ (nearly constant), and angles, $x$ (nearly rigidly rotating with frequency $\Omega(y)$). We study angle-action maps that are close to symplectic and have a positive-definite twist, the derivative of the frequency map, $D\Omega(y)$. When the map is symplectic, Nekhoroshev's theorem implies that the actions are confined for exponentially long times: the drift is exponentially small and numerically appears to be diffusive. We show that when the symplectic condition is relaxed, but the map is still volume-preserving, the actions can have a strong drift along resonance channels. Averaging theory is used to compute the drift for the case of rank-$r$ resonances. A comparison with computations for a generalized Froeschl\'e map in four-dimensions, shows that this theory gives accurate results for the rank-one case.  
\end{abstract}


\section{INTRODUCTION}\label{sec:Intro}
Arnold noted in his 1964 famous paper, ``Instability of Dynamical Systems with Several Degrees of Freedom", that the stability of an elliptical equilibrium in a Hamiltonian system with three or more degrees of freedom is not guaranteed \cite{Arnold64}. Of course transport in typical, nearly-integrable Hamiltonian systems, and correspondingly in symplectic maps, is inhibited because---by KAM theory---much of the phase space is foliated by invariant tori on which the dynamics is quasiperiodic \cite{Arnold63, Poshel01, delaLlave01}. In addition, Nekhoroshev showed that in certain cases---as we review in \Sec{Nekhoroshev}---the transport in chaotic regions of phase space is also small on an exponentially long time scale \cite{Nekhoroshev77, Lochak92}, implying a ``quasi-stability" of the dynamics, i.e., that the ``Arnold diffusion" is exponentially small. We will show that this quasi-stability can dramatically fail for a volume-preserving map that is ``nearly" symplectic.

In this paper we will study transport in maps $(x',y') = f(x,y;\eps)$ on $\bT^d\times \bR^k$, with $d$ angle, and $k$ action variables that are volume-preserving but need not be symplectic. One well-studied example of such a map is a volume-preserving discretization of the Arnold-Beltrami-Childress flow \cite{Arnold65}, called the ``ABC map" \cite{Cartwright94}; if two parameters of this three-dimensional map are small, it can be thought of as having two angles and one action.

We assume that the map $f$ has a formal parameter $\eps$, such that when $\eps = 0$ the actions are constant and the angles undergo a rigid rotation, with an action-dependent rotation vector. It is convenient to write such a map in the form 
\bsplit{ActionDepMap}
	x' &= x + \Omega(y') + \eps X(x,y;\eps) ,\\
	y' &= y + \eps Y(x,y;\eps) .
\esplit
Here $\Omega: \bR^k \to \bR^d$, the \textit{frequency map}, is evaluated at the image point $y'(x,y)$, and $X: \bT^d\times \bR^k \to \bT^d$ and $Y :\bT^d\times \bR^k \to \bR^k$ are perturbing ``forces".
For the simplest case we set $X \equiv 0$ and $Y(x,y;\eps) = F(x)$ to be independent of the actions, giving
\bsplit{GenFroMap}
	x' &= x + \Omega(y') ,\\
	y' &= y + \eps F(x) .
\esplit
This form is volume-preserving for any smooth $\Omega$ and $F$, see \Sec{Nekhoroshev}. It generalizes Chirikov's area-preserving map (where $d=k=1$) \cite{Chirikov79a}, Froeschl\'e's symplectic map (where $d=k=2$) \cite{Froeschle72}, and the volume-preserving normal form for a rank-one resonance ($d \ge 2$, $k=1$) \cite{Dullin12a}. 

The maps \Eq{ActionDepMap} and \Eq{GenFroMap} are integrable whenever $\eps=0$ in the sense that the actions are constant so that each $d$-torus $\cT_y = \{(x,y)\,|\, x \in \bT^d\}$ is invariant. 
In this case the angles evolve by rigid rotation with the rotation vector $\Omega(y)$. 
Denoting orbits of the map by sequences $(x^{(t)},y^{(t)}) = f^{t}(x^{(0)},y^{(0)})$, the dynamics of the integrable case are
\beq{IntegrableOrbit}
	(x^{(t)},y^{(t)}) = (x^{(0)} + \omega t,\,\, y^{(0)}),
\eeq
with the fixed rotation vector $\omega = \Omega(y^{(0)})$. When $\omega$ is incommensurate, the orbit is dense on the $d$-torus, $\cT_{y^{(0)}}$.

A prominent feature of the dynamics of \Eq{GenFroMap} are resonances. 
Generalizing \Eq{IntegrableOrbit}, the rotation vector $\omega \in \bR^d$ of an orbit $\{(x^{(t)},y^{(t)}): t \in \bZ\}$ is obtained by lifting the angle variable to $\bR^d$ and computing
\[
	\omega(x^{(0)},y^{(0)}) = \lim_{t\to\infty} \frac{1}{t}(x^{(t)}-x^{(0)}),
\]
if the limit exists. The orbit is said to be in a resonance if $\omega$ satisfies at least one relation of the form $m \cdot \omega = n$ for $m \in\bZ^{d}\setminus\{0\}$ and $n \in \bZ$. 
For a given $\omega$, the set of all vectors $m$ that satisfy such a relation,
\beq{resModule}
	\cL(\omega) = \{ m \in \bZ^d : m \cdot \omega \in \bZ\}, 
\eeq
is called the \emph{resonance module}.
For an \emph{incommensurate} frequency $\cL(\omega) = \{0\}$. The invariant tori of KAM theory are strongly incommensurate---they satisfy a Diophantine condition. If $\dim{\cL(\omega)} > 0$, then the frequency is \emph{commensurate}. The \emph{rank} of a resonance is the dimension of $\cL(\omega)$, the number of independent resonant $m$ vectors. Of course when $\eps = 0$ for \Eq{GenFroMap}, $\omega = \Omega(y)$, so every initial condition on the sets $m \cdot \Omega(y) = n$ is resonant. The existence of resonances has a profound effect on the dynamics.

Though it is clear that when $\eps \ll 1$ the actions evolve slowly under \Eq{GenFroMap}, there can be stronger constraints on the action dynamics for long or for infinite times. One such constraint is due to KAM theory \cite{Poshel01, delaLlave01}. A version of the KAM theorem has been proven for real analytic maps of the form \Eq{GenFroMap} with one action ($k=1$) in \cite{Cheng90a, Xia92} and for multiple actions in \cite{Cong96}. These theorems imply, under a R\"ussmann-type nondegeneracy condition on $\Omega$ and an intersection property on $F$, that there is a Cantor set of invariant tori when $\eps \ll 1$. The intersection property is the requirement that every invariant $d$-torus sufficiently close to the constant torus $\cT_y$ intersects its image. When $k = 1$, the codimension-one KAM tori constrain transport: no trajectories can cross such a torus.

However when $k >1$, the existence of codimension-$k$ tori no longer implies eternal stability, leading to the famous Arnold instability mechanism \cite{Arnold64}.
Nevertheless when a map the form \Eq{ActionDepMap} is symplectic, Nekhoroshev's theorem, which we recall in \Sec{Nekhoroshev}, can imply long time stability. In a symplectic map, the angle and action variables are canonically paired, so one must have $k=d$.
Moreover, as we discuss in \Sec{Nekhoroshev}, when \Eq{GenFroMap} is (canonically, exact) symplectic, the force must be conservative:
\beq{Gradient}
	F(x) = -\nabla V(x),
\eeq
for some potential $V: \bT^d \to \bR$, and the frequency must be generated by a function $S: \bR^d \to \bR$:
\[
	 \Omega = \nabla S(y).
\]
It is most common to assume that $S(y) = \tfrac12 \|y\|^2$, so
that $\Omega$ is the identity map. There have been many studies of transport in the symplectic case ranging from early studies \cite{Kaneko85, Kantz88, Kook90}, to more recent results \cite{Honjo03, Lega03, Honjo05, Guzzo05, Froeschle06, Guzzo11, Efthymiopoulos13, Cincotta14}. 

In this paper we we will compare the dynamics of symplectic and volume-preserving maps of the form \Eq{GenFroMap} for the four-dimensional case, $k=d=2$, with
\bsplit{GFForce}
	\Omega(y) &= y ,\\
	F(x)      & = -\frac{1}{2\pi}\left[\begin{array}{c} a\sin (2\pi x_1) + c \sin(2\pi(x_1+x_2)) \\ 
	             b\sin (2\pi x_2) + c \sin(2\pi(x_1+x_2 + \vphi)) \end{array}\right] .
\esplit
Here the terms with amplitudes $a$, $b$, and $c$ represent forcing of the $(1,0,n)$, $(0,1,n)$, and $(1,1,n)$ resonances respectively. 
Note that $F$ is a gradient only when the phase $\vphi = 0 \mod 1$. This case is Froeschl\'e's symplectic map \cite{Froeschle72}, and Nekhoroshev's theorem (see \Th{Nekhoroshev}) applies.
The case that $\vphi = \tfrac12$ and $b = 0$ in \Eq{GFForce} has also been studied previously, beginning with the work of Pfenniger \cite{Pfenniger85b} on the Hamiltonian-Hopf bifurcation, see also \cite{Jorba04, Zachilas13}. Though the force  for $\vphi = \tfrac12$ is not a gradient, the map becomes symplectic under a noncanonical coordinate transformation of the form $(x_1,x_2,y_1,y_2) \to (x_1,x_2,y_1,-y_2)$ (however, this makes the twist indefinite, see \Sec{Nekhoroshev}). We will use $\vphi$ as a convenient parameter to measure the deviation from symplecticity.

At the simplest level, Nekhoroshev's quasi-stability for \Eq{GenFroMap} is related to the fact that when an orbit is in an $m \cdot \Omega(y) = n$ resonance, the most important terms in the Fourier expansion of the force are those that have the angle dependence $m \cdot x$, since these vary slowly. When the force is a gradient \Eq{Gradient}, then the resonant Fourier terms in $F$ are of the form
\[
	\tilde{F}_m \propto m \sin(2 \pi m \cdot x + \theta) , 
\]
that is, they are in the $m$ direction. However, if $\Omega(y) = y$, then $m$ is orthogonal to the resonance manifold, $m \cdot y = n$, and the dynamics induced by the resonant force will be small, pendulum-like oscillations transverse to the resonance. The more profound result of Nekhoroshev is that the motion along resonance manifolds is in fact exponentially slow. By contrast, when $F$ is not a gradient, the resonant force can be in any direction, and in particular can have components along the resonance manifold, causing strong drifts, as we will see in \Sec{Numerics}. We will extend the argument above to more general frequency maps in \Sec{ResonantCoordinates}, then apply first order averaging theory to compute these drifts in \Sec{RankOneDrift}, and finally show that this calculation agrees with numerical simulations in \Sec{FroDrift}.

\section{SYMPLECTIC AND VOLUME-PRESERVING MAPS}\label{sec:Nekhoroshev}

Though the angle-action map \Eq{GenFroMap} is not always symplectic, it always preserves volume: there is a volume form $\vol$ such that $f^*\vol = \vol$. Indeed for any smooth frequency map and force, \Eq{GenFroMap} preserves the standard volume,
\beq{VolumeForm}
	\vol = dx_1 \wedge dx_2 \wedge \ldots \wedge dx_d \wedge dy_1 \wedge \ldots \wedge dy_k ,
\eeq
so that volume preservation is equivalent to $\det Df = 1$, where $Df$ is the Jacobian. 
This is easiest to see by writing the map as the composition $f = f_2\circ f_1$ where
\[
	f_1(x,y) = (x,y + \eps F(x))\qquad\mbox{and}\qquad f_2(x,y) = (x + \Omega(y),y)
\]
are shears. Much of the study of the dynamics of volume-preserving maps has concentrated on three-dimensional maps $k+d=3$, especially for ``one-action" ($k=1$) and ``two-action" ($k=2$) \cite{Cartwright94, Cartwright96, Mezic01} cases.

For the most of this paper, we will assume that $d=k$: the number of angle and action variables are the same and
$(x,y) \in \bT^d \times \bR^d$. This is especially appropriate for symplectic maps where each angle variable $x_i$ is canonically paired with an action variable $y_i$. In the canonical case the map preserves the two form
$\sigma = dy \wedge dx = \sum_{i=1}^d dy_i \wedge dx_i$; that is, $f$ is symplectic when $f^* \sigma = \sigma$ \cite{Arnold78}, or
\beq{Poisson}
	Df^T J Df = J, \qquad J = \begin{pmatrix} 0 & -I \\ I & 0 \end{pmatrix} ,
\eeq
where $J$ the $2d \times 2d$ Poisson matrix.\footnote
{More generally, $\sigma$ is only required to be a nondegenerate two-form, though Darboux's theorem implies that there are local canonical coordinates \cite{Arnold78}.}
Every such symplectic map preserves the $2d$-volume form \Eq{VolumeForm}.

The generalized \fro map \Eq{GenFroMap} has Jacobian 
\beq{Df}
	Df(x,y) = \begin{pmatrix} I + \eps D\Omega(y + \eps F(x)) DF(x) & D\Omega(y +\eps F(x)) \\
						  \eps DF(x) & I
		 \end{pmatrix} .
\eeq
This matrix satisfies \Eq{Poisson} only when the $d\times d$ matrices $DF$ and $D\Omega$ are symmetric. Indeed, setting the formal parameter $\eps = 1$ and dropping the arguments from $D\Omega(y')$ and $DF(x)$, gives
\[
	Df^T J Df = \begin{pmatrix} DF^T-DF + DF^T(D\Omega-D\Omega^T) & -I + DF^T(D\Omega-D\Omega^T) \\
				I +(D\Omega-D\Omega^T)DF & D\Omega -D\Omega^T \end{pmatrix}.
\]
In order that this reduce to $J$, the lower right block must vanish, so $D\Omega$ is symmetric, and then vanishing of the upper left block implies that $DF$ is also symmetric.

A special category of symplectic maps are the \emph{exact} maps. The symplectic form $\sigma$ on $\bT^d \times \bR^d$ is exact: there is a Liouville one-form $\nu = y \cdot dx$, so that $\sigma = d\nu$.
A map $f$ is exact symplectic with respect to $\nu$ if there exists a discrete \textit{Lagrangian} $L(x,y,x',y')$ such that
\beq{exactSymplectic}
	f^* \nu - \nu = y'\cdot dx' - y \cdot dx = dL ,
\eeq
on the graph $(x',y') = f(x,y)$. Of course every such map is also symplectic since 
$d (f^*\nu) - d\nu = f^*\sigma -\sigma = d^2L = 0$.
In order that the map be well-defined on the torus, $L(x+ m,y,x'+m,y') = L(x,y,x',y')$ for each $m \in \bZ^d$. For example, if
\beq{Lagrangian}
	L(x,y,x',y') = (x'-x)\cdot y' - S(y') - \eps V(x),
\eeq
for a periodic potential $V$, then rearranging \Eq{exactSymplectic} gives $x' \cdot dy' + y \cdot dx = d( x \cdot y' +S(y')+ \eps V(x) )$, or equivalently
\bsplit{symMap}
	x' &= x + \nabla S(y') ,\\
	y' &= y - \eps \nabla V(x) .
\esplit
This is the generalized \fro map \Eq{GenFroMap} with the force $F(x) = -\nabla V(x)$ and frequency map $\Omega(y) = \nabla S(y)$. Of course in this case $DF$ and $D\Omega$ are automatically symmetric.

When $\eps = 0$, the map \Eq{symMap} is integrable in the sense of Liouville: the $d$-action variables are invariants. As we noted in \Sec{Intro}, the dynamics on each $d$-dimensional torus corresponds to rigid rotation
with rotation vector $\omega = \Omega(y)$. The \emph{twist} of the map is the symmetric matrix $D\Omega = D^2S$, and the map is said to have \textit{positive-definite twist} on a set $B \in \bR^d$ when $D\Omega$ is uniformly positive definite; i.e., when there exists a $C >0$ such that
\beq{posDefinite}
	v^T D\Omega(y) v \ge C \|v\|^2, \quad \forall y \in B.
\eeq
In this case the frequency  varies nontrivially from torus-to-torus. 

In certain cases, the orbits of \Eq{symMap} do not stray too far from this integrable case when $\eps \ll 1$. In particular for the symplectic case, Nekhoroshev's theorem asserts that the actions of \textit{all} orbits do not drift far over exponentially long times. Although originally proven for Hamiltonian flows \cite{Nekhoroshev77}, Nekhoroshev's theorem applies to exact symplectic maps by flow interpolation \cite{Kuksin94}. It was also proven directly for the positive-definite twist case by Guzzo:

\begin{teo}{\cite{Guzzo04}}\label{thm:Nekhoroshev}
Suppose that $B \subset \bR^d$ is an open set, $S: B \to \bR$, and $V: \bT^d \to \bR$ are analytic, and $D\Omega = D^2S$ is uniformly positive definite on $B$. Then for orbits of \Eq{symMap} there exist positive constants $\eps_0,\alpha,\beta,T_0,C_0$ such that for any $\eps < \eps_0$, and for any $(x^{(0)},y^{(0)})\in \bT^d \times B'$, with $B' = \{y\in B:\text{dist}(y,\partial B)>2C_0\eps^\alpha\}$,
\[
	\|y^{(t)} - y^{(0)}\|\leq C_0\eps^\alpha
\]
for any $t\in\bZ$ satisfying
\[
	|t|\leq T_0\exp\left(\frac{\eps_0}\eps\right)^\beta .
\]
\end{teo}

\noindent
Informally this theorem implies that the drift in the action variables is very slow. The exponents obtained vary, depending upon the methods used to prove the theorem, but typically $\alpha \le \tfrac12$---the upper limit being determined by the width of typical resonances, and $\beta \simeq(2d)^{-1}$.

In the next sections, we will contrast the action transport for the map \Eq{GenFroMap} when it is constrained by \Th{Nekhoroshev}, with the action transport when the map is only volume preserving but still satisfies the remaining hypotheses of \Th{Nekhoroshev}.

\section{VISUALIZING TRANSPORT AND DRIFT}\label{sec:Numerics}

The visualization of the dynamics of a four or higher-dimensional map is a challenge. A key numerical tool is to compute some scalar quantity on strategically chosen two-dimensional slices of initial conditions. To delineate chaotic regions one would like to compute the largest Lyapunov exponent; however, it is difficult to obtain accurate results. A more well-behaved quantity is the Fast Lyapunov Indicator (FLI) \cite{Froeschle97, Froeschle06},
\beq{FLI}
	\gamma(x^{(0)},y^{(0)};v,T) = \sup _{\tau\leq T}\,\log \left\|\prod_{t=0}^{\tau-1}Df(x^{(t)},y^{(t)})v\right\|.
\eeq
where $v$ is some fixed deviation vector. The use of the supremum allows the FLI to distinguish orbits using a smaller number of iterations, $T$, than needed for the Lyapunov exponent to converge. As a first illustration consider the decoupled case $c=0$ for \Eq{GFForce}. In each canonical plane the map is then an area-preserving Chirikov map. The FLI for the $(x_1,y_1)$ plane with $\eps a = 0.52$ is shown in \Fig{stdmapfli}. The color bar shows the range $1 \le \gamma \le 10$ for the FLI \Eq{FLI}. The highest FLI values occur in the chaotic separatrix layer surrounding the $(0,1)$ resonance. As noted in \cite{Guzzo02} the FLI also distinguishes between resonant (dark blue) and rotational (KAM) invariant (light blue) circles.

\InsertFig{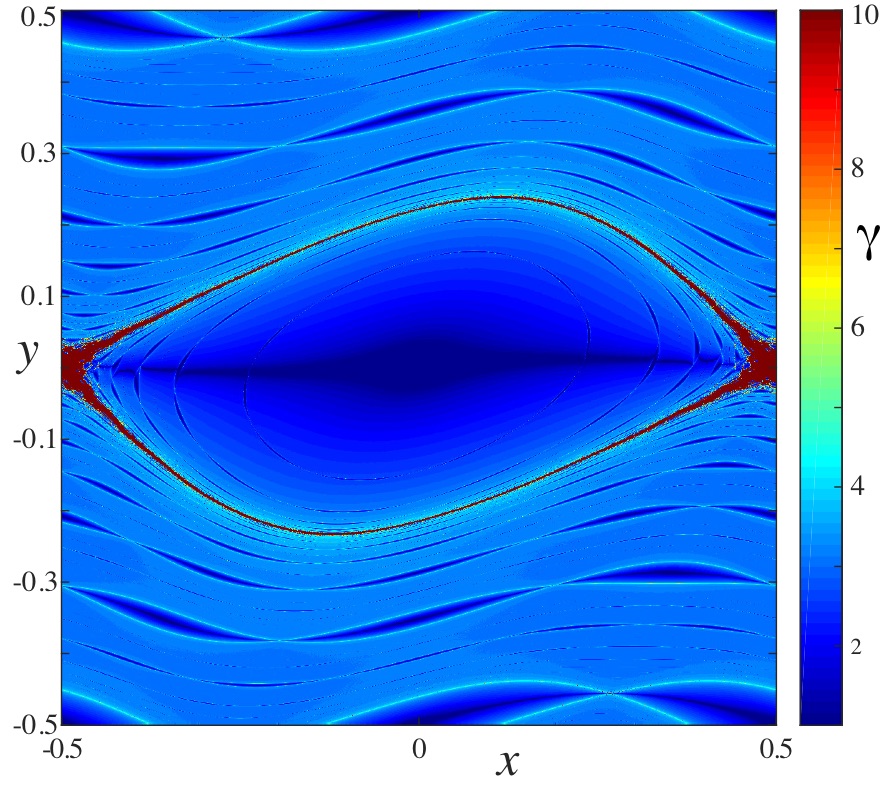}{Fast Lyapunov indicator (FLI) as a function of initial condition for the 2D Chirikov map with $\eps a=0.52$, with $10^6$ initial conditions on a square grid, each iterated $T= 10^3$ steps. The bar (color online) indicates the values of \Eq{FLI} for $v = ((3-\sqrt{5})/2,1)$.}{stdmapfli}{3.5in}

We can similarly visualize the structure of the phase space in four-dimensions by simultaneously plotting the FLI on several complementary planes of initial conditions \cite{Lega16}.
We show three planes for the symplectic \fro map, assembled into faces of a cube, in \Fig{FroeschleFLICube}. The two vertical planes are canonical: the left face is the two-plane $(x_1,0.0,y_1,0.65)$ and the right face is the two-plane $(0,x_2,0.65,y_2)$. Each of these canonical planes has resonant structures resembling the 2D case of \Fig{stdmapfli}, though the amplitudes of the $(0,1,0)$ and $(1,0,0)$ resonances are smaller here since $\eps a = \eps b = 0.1$. A new feature is a large resonant island that appears in each canonical plane near $y_i = 0.35$. This corresponds to the $(1,1,1)$ resonance driven by the coupling with amplitude $\eps c = 0.07$ in \Eq{GFForce}. For example since $y_2 = 0.65$ on the $(x_1,y_1)$ plane, the zero-order resonant condition $y_1+y_2 = 1$ corresponds to $y_1 = 0.35$. In the figure, the two canonical planes appear to be artificially joined along a ``line"---this is really the two-plane $y_1 = y_2 = 0.65$.

\InsertFig{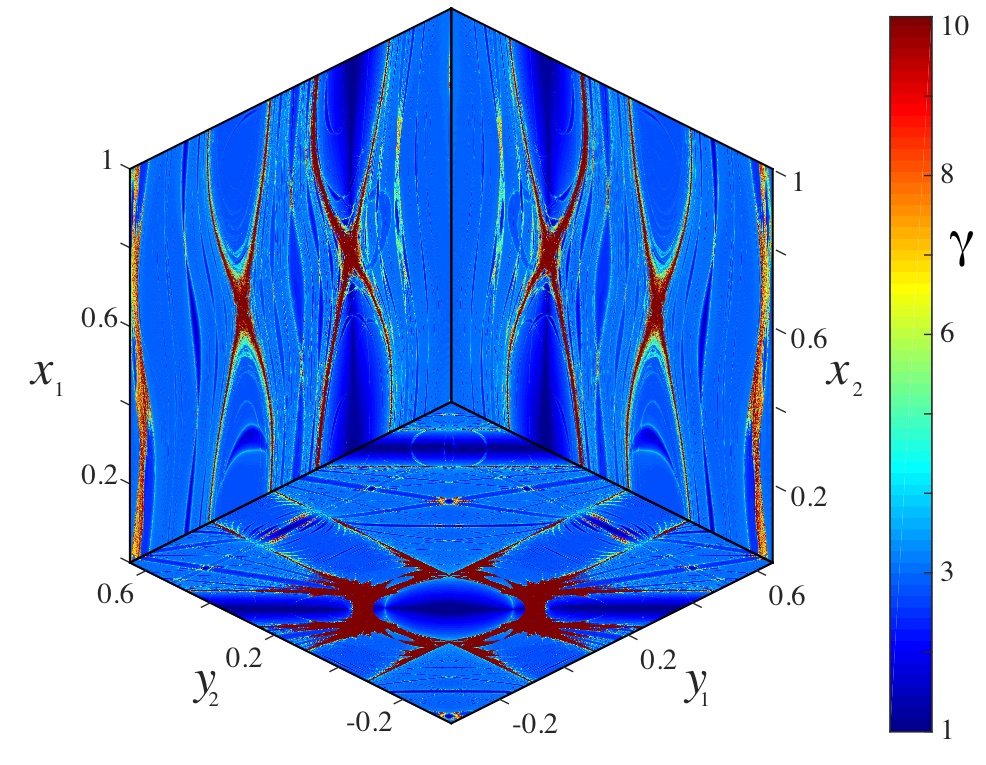}{Three FLI planes for the symplectic ($\vphi = 0$) \fro map with \Eq{GFForce} for $\eps (a,b,c) = (0.1,0.1,0.07)$: an action plane with fixed angles $x_1=x_2=0$, and two canonical planes with the other angle-action pair fixed at $(x_j,y_j) = (0,0.65)$, with $10^6$ initial conditions on each square grid, each iterated $T = 10^3$ steps using $v =((3-\sqrt{5})/2,1,1,1)$.}{FroeschleFLICube}{5in}

The third, horizontal two-plane in \Fig{FroeschleFLICube} is the action plane $(0,0,y_1,y_2)$. It is shown intersecting each canonical plane along the appropriate fixed-angle lines. In the action plane, each rank-one resonance becomes a strip centered on $m \cdot y = n$ with a width proportional to the square root of the amplitude of the resonant term in the force. In this plane, we see prominent strips for the $(1,0,0)$, $(0,1,0)$, $(1,1,0)$ and $(1,1,1)$ resonances, as well as a number of smaller resonant strips caused by nonlinear beating. Resonance overlap near the rank-two crossings of these resonances causes stronger chaotic regions. Such overlap necessarily occurs even for very small perturbations, and can cause drift along, and transfer between, resonances. It is the speed of this drift that is addressed by \Th{Nekhoroshev}.

These FLI cross-sections provide a look at the global structure of phase space, and can serve as a backdrop for orbit projections. For the symplectic case, orbits within or near a rank-one resonance oscillate orthogonally to the resonance, and \Th{Nekhoroshev} ensures any $\cO(1)$ drift parallel to resonances must occur over exponentially long times. Numerical studies have found that this slow process is diffusive \cite{Guzzo11} (see also \Fig{ActionDrift} below). 

No such long-time stability appears to exist for volume-preserving, nonsymplectic maps. Indeed, when the phase shift $\vphi$ is nonzero for the generalized \fro map, orbits in the $(1,1,n)$ resonances drift rapidly in action space, moving an $\cO(1)$ distance in a few hundred iterates. A first indication of this drift is visible in the FLI planes shown in \Fig{VPFroeschleFLICube}. In this figure the parameters $(a,b,c)$ are half the size of those in \Fig{FroeschleFLICube}; nevertheless, when $\vphi \neq 0$, the much of the resonance structure seen in the symplectic case is still present, However, the FLI plots indicate much stronger ``chaotic zones", where $\gamma \simeq 6$, in the $(1,1,0)$ and $(1,1,1)$ resonances. One must be cautious in attributing this to local structure in the map: the actions in these resonances drift far from their initial values.

\InsertFigTwo{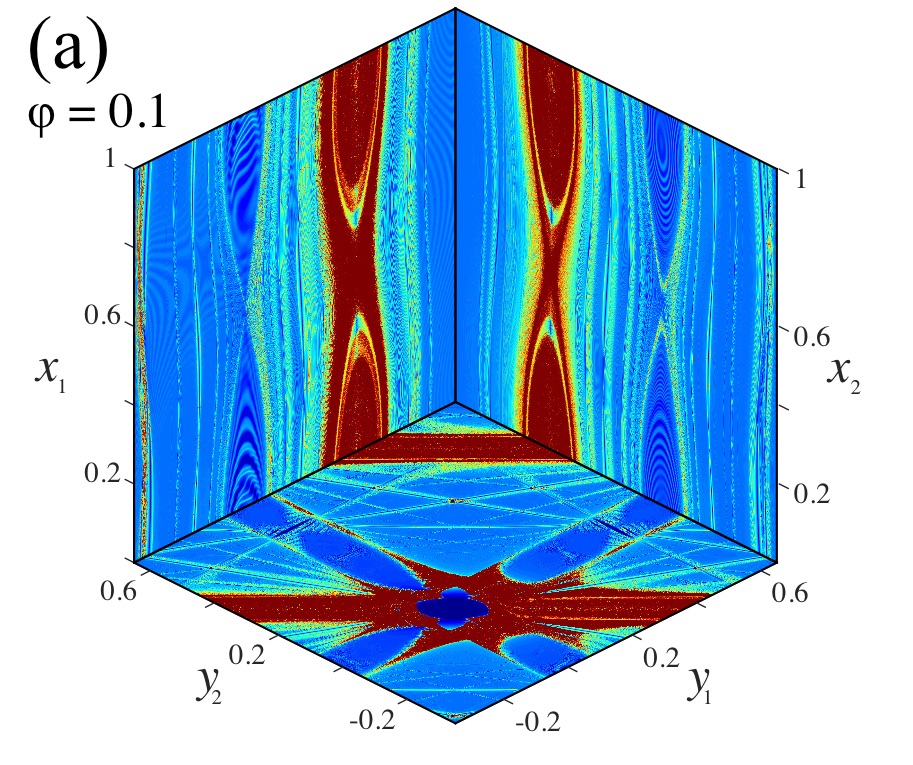}{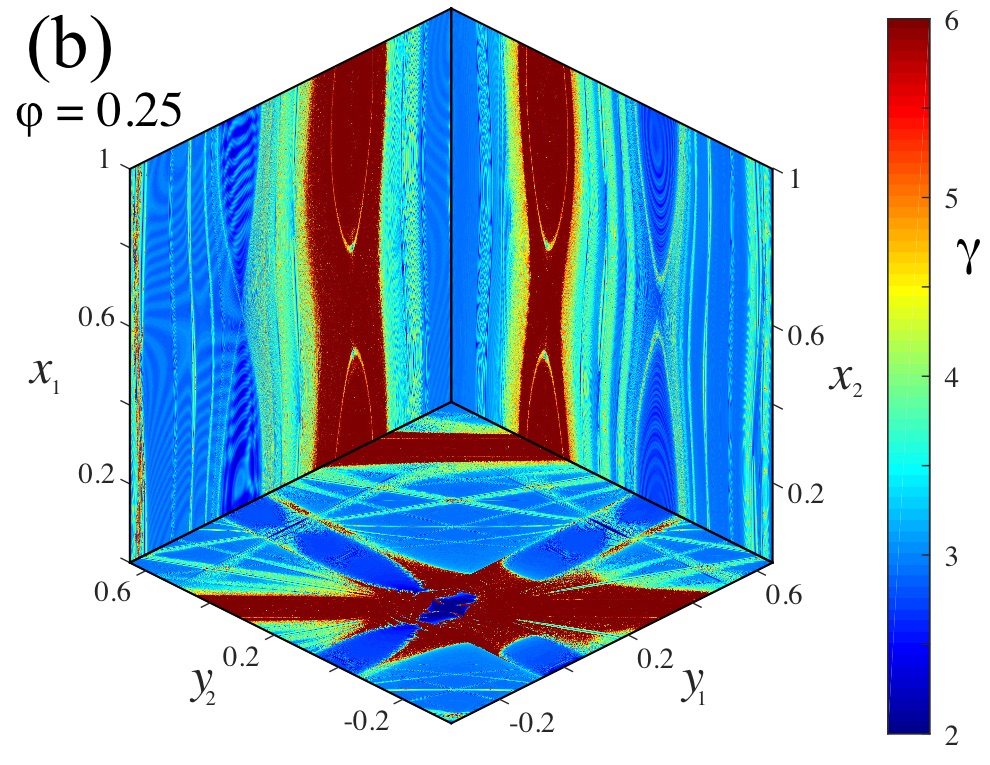}
{FLI planes (as in \Fig{FroeschleFLICube}) for the volume-preserving generalized \fro map with $\eps(a,b,c) = (0.05,0.05,0.035)$ and $\vphi = 0.1$ (a) and $0.25$ (b). The $10^6$ initial conditions on each plane are each iterated $10^3$ times.}
{VPFroeschleFLICube}{2.8in}

Indeed, one can visualize the drift on the action plane $(0,0,y_1,y_2)$ by looking at some representative orbits In \Fig{OrbitProjections}, the FLI is shown as a grayscale background, with black for $\gamma = 0$ and white for $\gamma = 6$. For the symplectic case (panel (a)) three orbits were initialized (small dots in green online) in the chaotic separatrix layer of three different resonances, $(1,1,0)$ (gold), $(0,1,0)$, (red) and $(1,0,0)$ (blue). Each initial condition was iterated $10^{10}$ times, and points are plotted only when $|x_1|,|x_2| < 0.001$. Note that the orbits are confined to a small interval of size $\cO(0.1)$  along the resonance and to a narrow chaotic zone that appears in the slice to be two sides of the separatrix layer of the resonance. 

By contrast, when $\vphi\neq 0$, the orbits drift along the resonances rapidly as shown in \Fig{OrbitProjections}(b). Here three initial conditions in the $(1,1,0)$ resonance are iterated only $200$ steps and \textit{all} points along the orbits are shown \textit{projected} onto the action-plane. Two of these orbits (red and blue) drift by $\cO(1)$ in this small number of iterations, and one (gold) appears to be, at least temporarily confined near $y_1 = -y_2 \approx -0.3$.

\InsertFigTwo{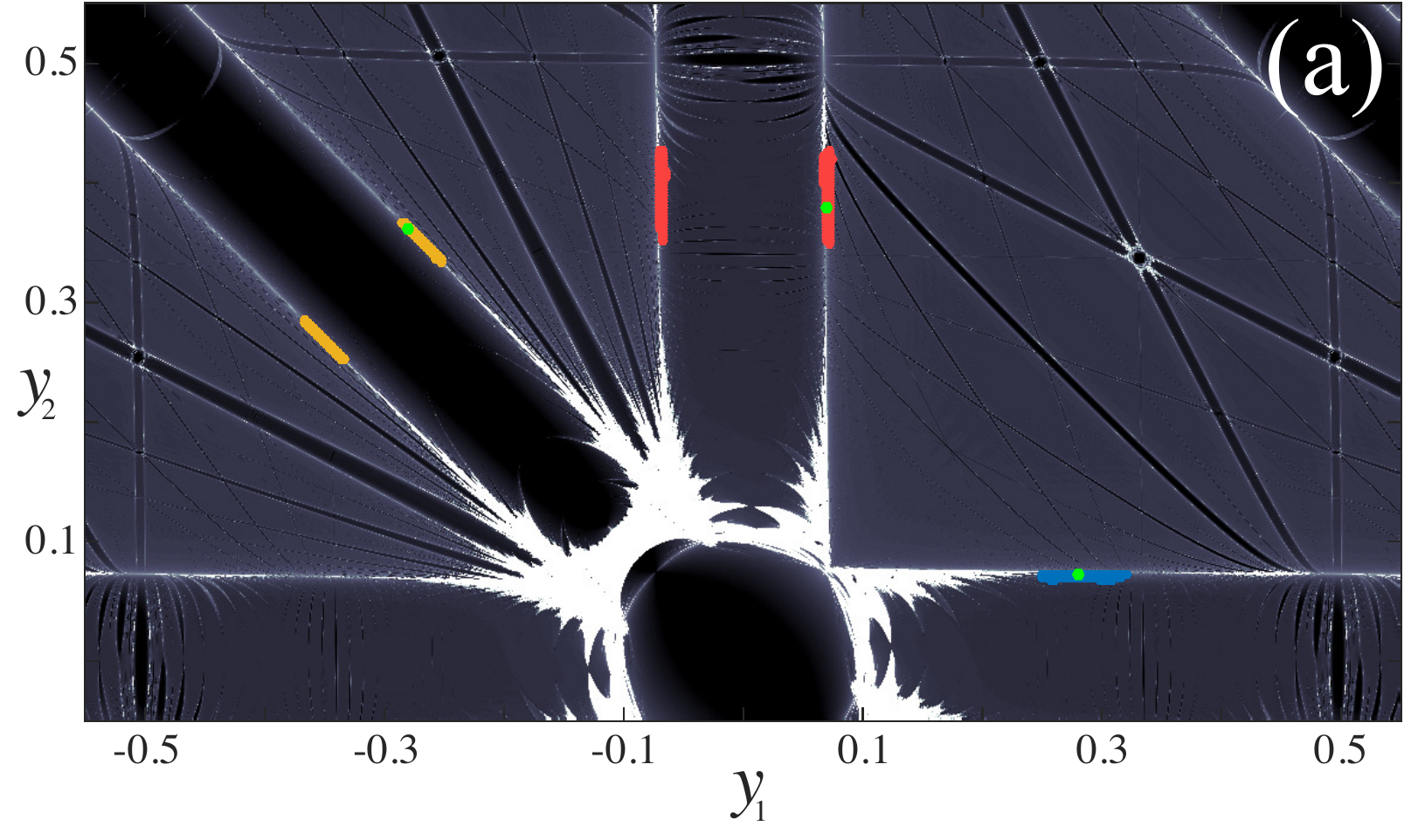}{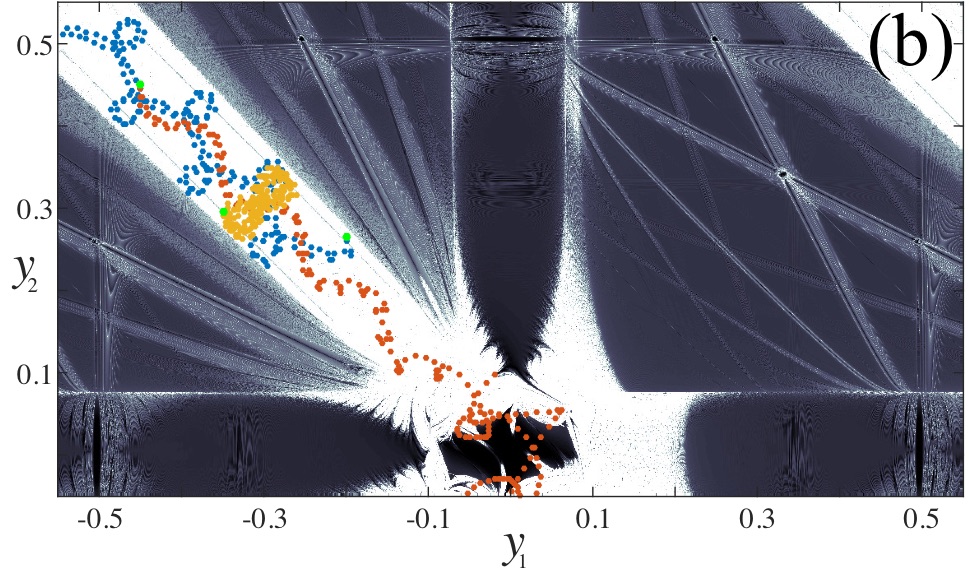}{FLI (grayscale background) for the \fro map with \Eq{GFForce} and $\eps (a,b,c) = (0.05,0.05,0.035)$ on the plane $(0,0,y_1,y_2)$. Three orbits (color online) are shown in each panel, starting from the small (green) dots. (a) Symplectic $(\vphi = 0)$ case with each orbit iterated $10^{10}$ steps, and only points in a slice near $(x_1,x_2) = 0$ are shown. (b) Volume-preserving $(\vphi = 0.25)$ case with each orbit iterated $200$ steps, and all points are shown projected onto the action-plane.}{OrbitProjections}{2in}

One way to better quantify the drift is to compute the action-diameter of an orbit
\beq{Diameter}
	D(x^{(0)},y^{(0)}; T) = \max_{0\leq t,s\leq T} \|y^{(t)} - y^{(s)} \|.
\eeq
A computation of the diameter as a function of initial condition in the action plane is shown in \Fig{ActionDiameter}. For the symplectic case, the largest diameters occur at the intersection of the $(1,1,1)$ and $(0,1,1)$ resonances, where there is overlap-induced chaos. When $\vphi \neq 0$, however, the diameter is nearly an order of magnitude larger almost everywhere in the $(1,1,1)$ resonance, except near the rank-two resonance $\omega = (0,0)$ which is apparently still foliated with invariant tori near the zero-angle plane, recall \Fig{OrbitProjections}.

\InsertFigTwo{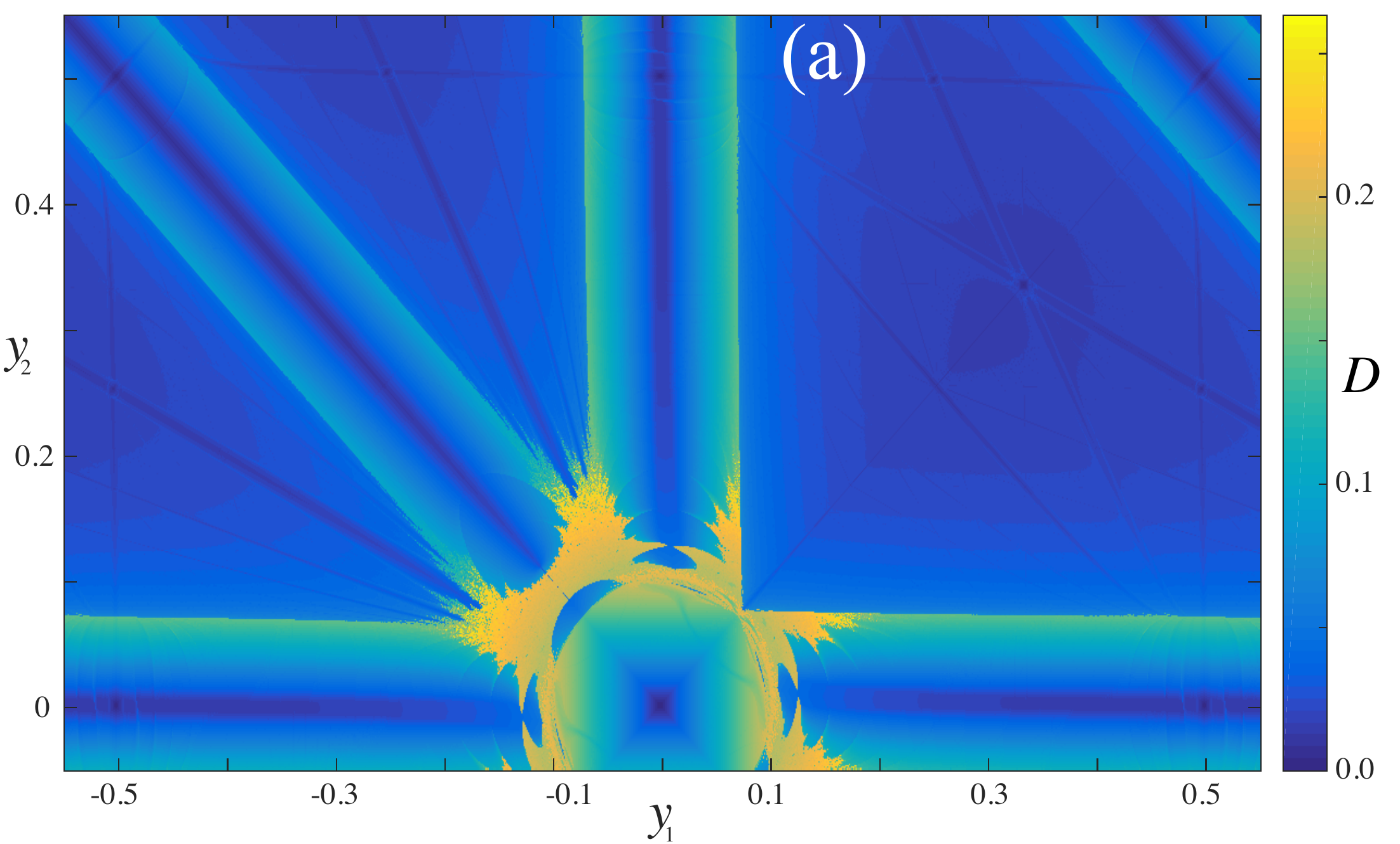}{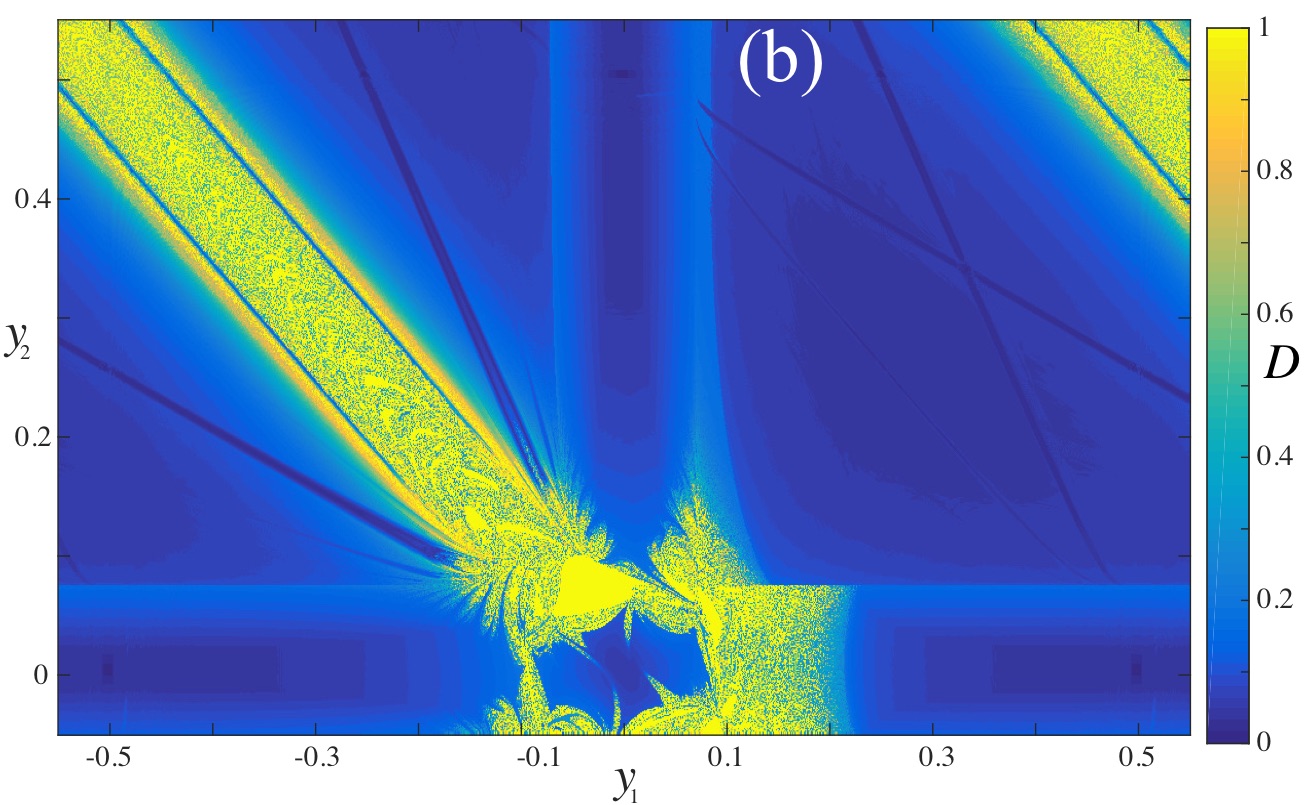}
{Orbit diameter \Eq{Diameter} for initial conditions in the $(0,0,y_1,y_2)$ plane for \Eq{GenFroMap} and \Eq{GFForce} with $\eps (a,b,c) = (0.05,0.05,0.035)$ for $T=10^3$ iterates. (a) The symplectic case, $\vphi = 0$. (b) The volume-preserving case, $\vphi = 0.25$. Note that the shading scales (color online) are different for the two panels.}{ActionDiameter}{2.2in}

The strong effect of nonzero $\vphi$ on the dynamics can also be seen by looking at the mean-square displacement of the actions. Figure~\ref{fig:ActionDrift} shows the square of the change in action, $y^{(t)}-y^{(0)}$. For panel (a), the symplectic case, this is averaged over a small ball of initial conditions in the separatrix layer of the $(1,0,0)$ resonance, and the $y_2$ component is shown to have a linearly increasing mean-square drift: the motion is diffusive. By contrast \Fig{ActionDrift}(b) shows that, for an initial condition in the $(1,1,0)$ resonance for the volume preserving case, the norm of the action has a squared drift that increases as $T^2$: the motion is ballistic.

\InsertFigTwo{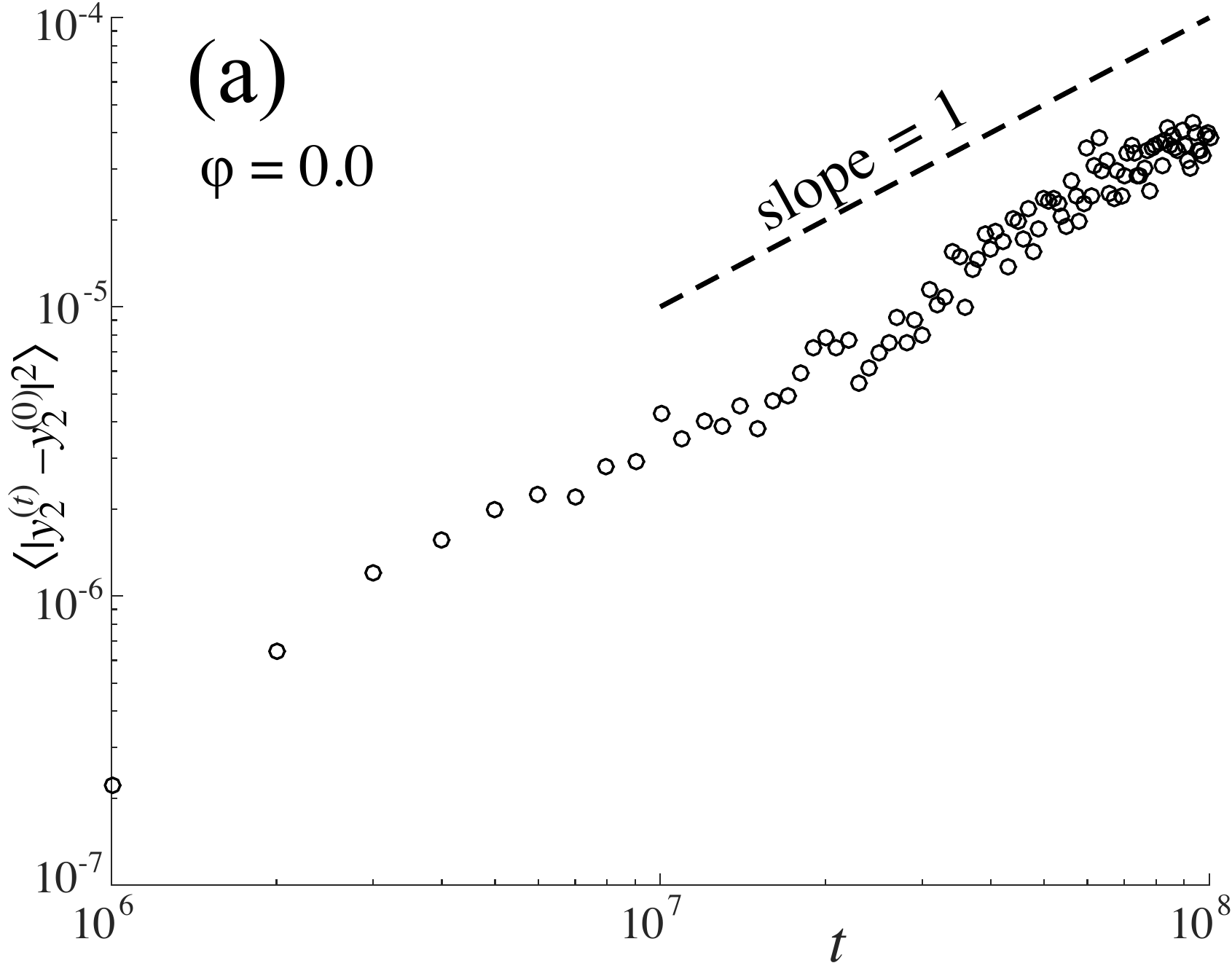}{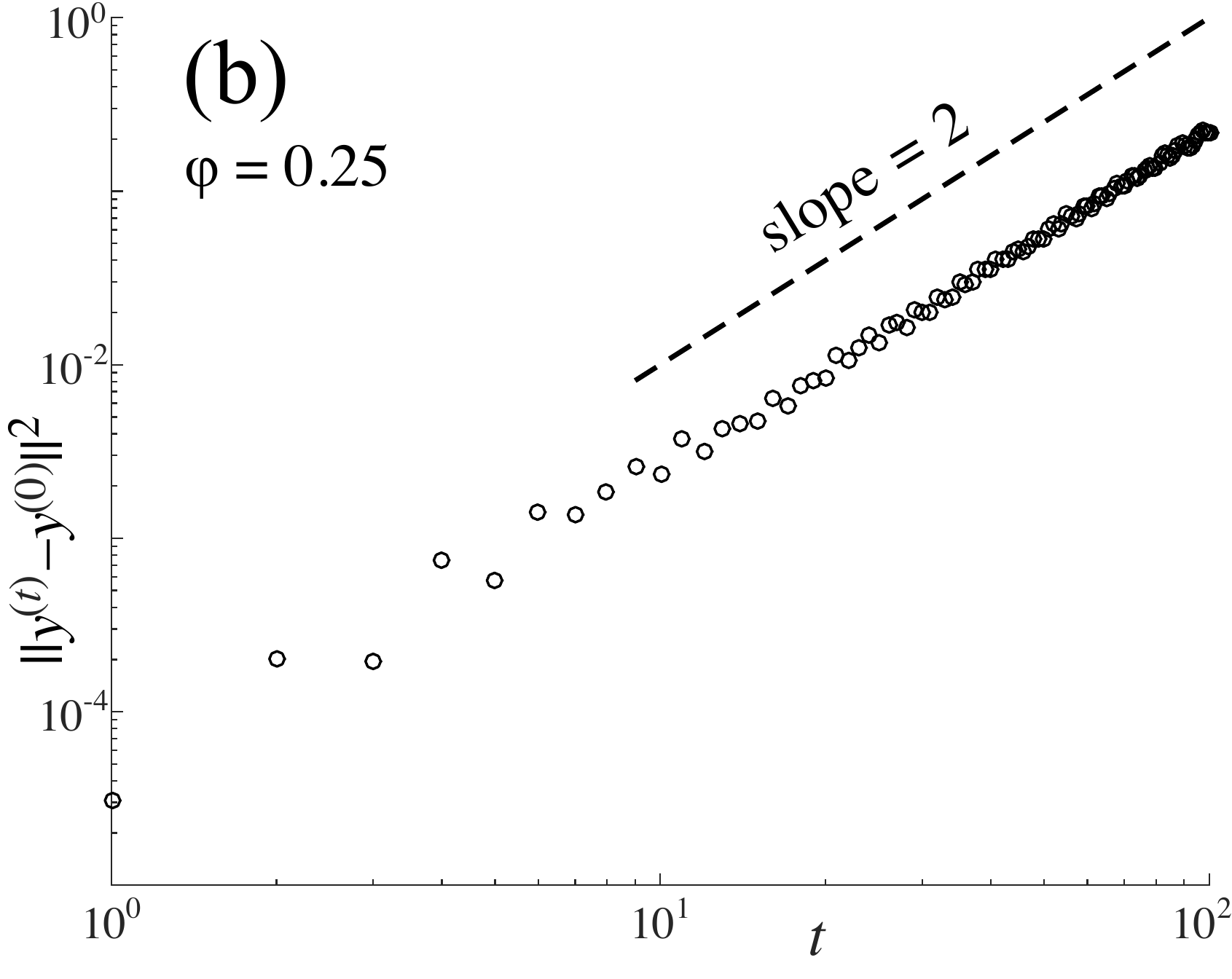}{Growth of the squared action displacement for the same parameters as \Fig{ActionDiameter}. (a) Symplectic case. Mean-squared displacement for 100 initial conditions within $10^{-3}$ of (0.5,0,0,0.35) is still less than $10^{-4}$ after $10^8$ iterations. (b) Volume-preserving case. Squared displacement nears $\cO(1)$ for initial condition (0,0,-0.45,0.45) after only 100 iterations.}{ActionDrift}{2.5in}

To better understand drift direction we want to compute an approximate action velocity. It is useful to compute a running average by applying \emph{exponential smoothing}. To compute a linear trend for the actions, we use the ``Holt-Winters" double exponential smoothing \cite{Gardner06}, see App B. Figure~\ref{fig:speedcubes} shows the the trend, $V^{(T)}$, from \Eq{HoltWinters} in the $(1,-1)$ direction for $T=100$ on three two-planes of initial conditions. The maximum action drift speeds in this direction are $\approx 5(10)^{-4}$ in the $(1,1,n)$ resonances. This speed is the same as that observed in \Fig{ActionDrift}(b), since there $(\Delta y)^2 \approx 0.25$ after $100$ iterates.

\InsertFigTwo{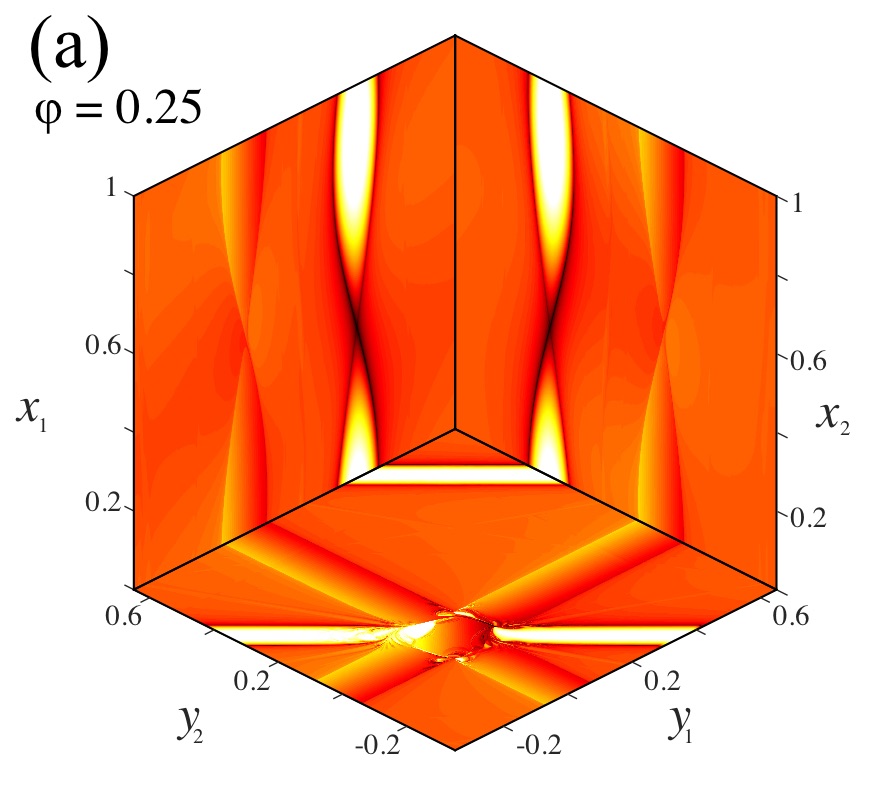}{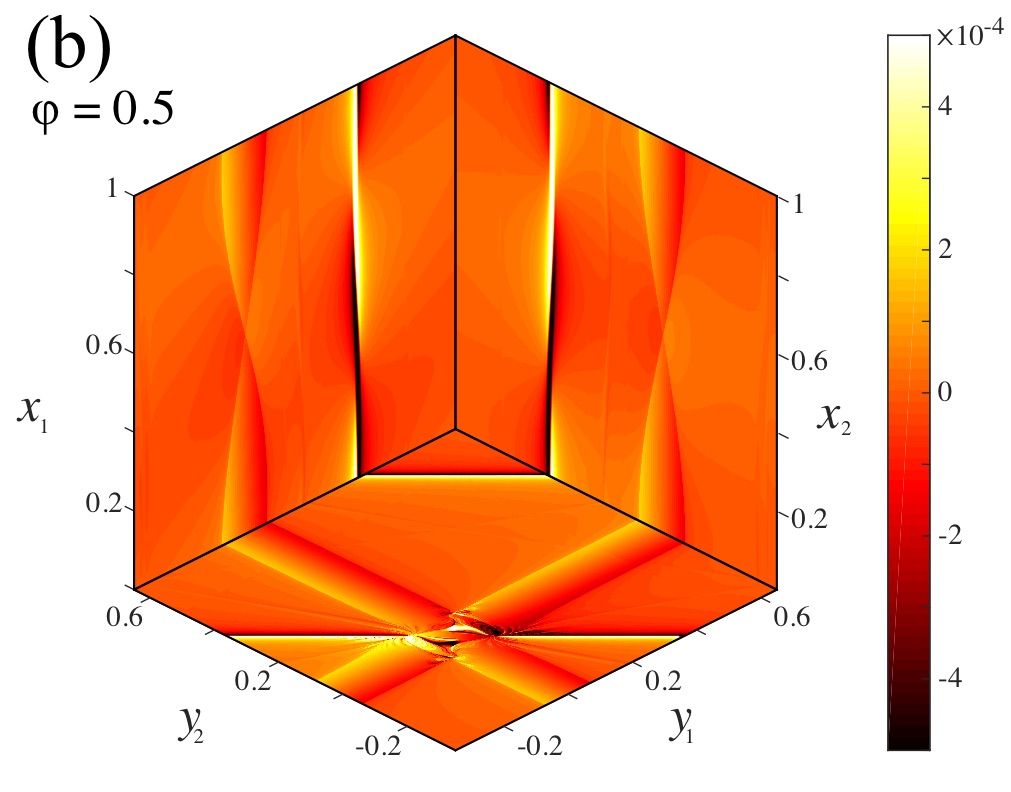}{Smoothed trend planes for the change in action in the (1,-1) direction, for the volume-preserving generalized \fro map for $\eps (a,b,c) = (0.05,0.05,0.035)$ for (a) $\vphi = 0.25$ and (b) $\vphi=0.5$. The shading scale (color online) represents $V^{(T)} \cdot (1,-1)/\sqrt{2}$ for $T=100$.}{speedcubes}{3in}

\section{RESONANT COORDINATES}\label{sec:ResonantCoordinates}

In this section we will study the drift in the action variables for a map of the form \Eq{ActionDepMap} for an initial condition in the neighborhood of a point in a rank-$r$ resonance. We will assume that the frequency map is symmetric and uniformly positive definite, \Eq{posDefinite}. 

We will show that, in the correct coordinates, the map can be averaged over the (fast) non-resonant angles to produce an approximate dynamics in semi-direct product form. The resonant angle and action variables evolve on this time scale independently of the non-resonant actions. The latter are forced by the resonant dynamics, and we identify the terms that are responsible for causing the drift that we observed in \Sec{Numerics}.

\subsection{Resonances}
For a rank-$r$ resonance, the module $\cL(\omega)$  of \Eq{resModule} has a basis $\{m^{(1)},\ldots,m^{(r)}\} \subset \bZ^d$ such that whenever $n_j = m^{(j)}\cdot\omega$ then the components of each vector $(m^{(j)},n_j)$ are coprime (however, the components of a vector $m^{(j)}$ need not be coprime). The resonance module is the range of the $d \times r$ matrix
\[
	M = \left[m^{(1)} \, m^{(2)}\, \cdots\, m^{(r)}\right]
\]
over $\bZ$. 
Every such matrix has a Smith normal form \cite{Dummit04}, 
\beq{SmithForm}
	M = PSW ,
\eeq
with unimodular matrices $P \in Sl(d, \bZ)$ and $W \in Sl(r, \bZ)$, and a $d\times r$ diagonal matrix $S =\text{diag}(s_1,\ldots,s_r)$. Here the $s_j\in\bN$ have the property that $s_j$ divides $s_l$ for all $j<l$. 
We will use the transpose of $d \times d$ matrix $P$ to define new angles, $\xi = P^T x$; the unimodularity of $P$ will preserve periodicity. Notice the span of the first $r$ columns of $P$ is $\cL(\omega)$, so the first $r$ components of $\xi$ represent the \emph{resonant angles}. The corresponding new frequencies are
\beq{newFreq}
	\hat{\omega}= P^T \omega .
\eeq
If we define a vector $n\in\bZ^r$ with components $n_j$, then $n = M^T\omega$, and
\[
	\hat{n} = W^{-T}n = W^{-T}(W^TS^TP^T) \omega = S^T\hat{\omega} \in \bZ^r .
\]
Consequently
\beq{OmegaHat}
 \hat{\omega}= (\hat\omega_S, \hat\omega_F) 
             = \left[ \frac{\hat{n}_1}{s_1}\, \ldots\, \frac{\hat{n}_r}{s_r}\,\,
                      \hat{\omega}_{r+1}\, \ldots\, \hat{\omega}_{d} \right]^T .
\eeq
Since, by assumption $\text{dim}(\cL(\omega)) = r$, the components of $\hat{\omega}_F$ are incommensurate. We think of $\hat{\omega}_S$ as representing ``slow frequencies" and $\hat{\omega}_F$, ``fast frequencies".

Given a frequency map $\Omega: \bR^d \to \bR^d$, and a vector $(m,n) \in \bZ^d \times \bZ$, the subset
\beq{resMan}
	\cR_{m,n} = \{ y \in \bR^d: m \cdot \Omega(y) = n\}
\eeq
is a \textit{resonance manifold}.
Whenever $D\Omega$ is nonsingular, a rank-one resonance is a codimension-one submanifold. 
Given a point $y_* \in \cR_{m,n}$, then a nearby point $y \in \cR_{m,n}$ if
\[
	0 = m\cdot \Omega(y) - n = m\cdot D\Omega_*(y-y_*) + \cO(y-y_*)^2 ,
\]
where $D\Omega_* = D\Omega(y_*)$. Therefore resonance manifold is locally perpendicular to every vector of the form $D\Omega_*^T m$ for each $m \in \cL(\omega)$. 
A point $y_*$ lies on a rank-$r$ resonance if it is in the intersection of $r$, independent rank-one resonance manifolds, i.e., if the resonance module \Eq{resModule} has dimension $r$ at $\omega = \Omega(y_*)$. This defines a codimension-$r$ submanifold
\[
	\cR_\omega = \bigcap_{m \in \cL(\omega)} \cR_{m, m\cdot \omega} .
\]

When $D\Omega_*$ is symmetric, positive definite, it induces an inner product
\beq{innerProduct}
	\langle v,w\rangle_{D\Omega_*} \equiv v^T D\Omega_* w ,
\eeq
and a corresponding norm
$
 \| \cdot \|_{D\Omega_*} 
$ 
This norm can be used to find an orthogonal basis for the module $\cL(\omega)$ that will be used to define new action variables.\footnote
{If $D\Omega$ is nonsingular, but not symmetric, we can use the polar decomposition to obtain
an appropriate norm, giving essentially the same results.}
Applying the Gram-Schmidt process to the columns of $P$ from \Eq{SmithForm} with respect to the inner product \Eq{innerProduct} gives
\beq{PQR}
	P= QR, \quad Q^TD\Omega_*Q = I ,
\eeq 
where $R$ is upper triangular and $Q$ is orthogonal with respect to \Eq{innerProduct}. The first $r$ columns of $Q$ and $P$ have the same span over $\bR$ which contains $\cL(\omega)$, so the remaining columns of $Q$ are orthogonal to $\cL(\omega)$, and therefore span the tangent space to $\cR_{\omega}$ at $y_*$. 

\subsection{Expanding and Averaging}\label{sec:Averaging}

For a point $(x,y)\in\bT^d\times\bR^d$, consider a family of maps $f$ of the form \Eq{ActionDepMap} where $\Omega$, $X$, and $Y$ are real analytic, and $D\Omega$ is symmetric and uniformly positive definite, \Eq{posDefinite}.
Suppose $y_* \in \cR_{\omega}$ for a rank-$r$ rotation vector $\omega = \Omega(y_*)$, and $D\Omega_* \equiv D\Omega(y_*)$. To investigate the possibility of action drift along resonances, we want to define coordinates in a neighborhood $y_*$ that are ``natural" to the resonance. First, to balance the size of the perturbing terms we define new action variables $\delta$ by
\[
	y = y_* + \kappa \delta , \mbox{ where } \kappa \equiv \sqrt{\eps} .
\]
Expanding \Eq{ActionDepMap} then gives
\bsplit{ExpandGeneral}
	x'		&= x + \omega + \kappa D\Omega_* \delta' + \cO(\kappa^2) \mod 1 , \\
	\delta' &= \delta + \kappa Y(x,y_*;0) + \cO(\kappa^2) .
\esplit

We then define coordinates $(\xi, \eta)$ adapted to the resonance using the matrix $P$ from \Eq{SmithForm} and
the orthogonal matrix $Q$ from \Eq{PQR},\footnote
{This transformation is not canonically symplectic unless by chance $Q = P$, i.e.,
$P$ is $D\Omega_*$ orthogonal. For a general frequency map, the transformation is also not volume preserving, but has constant Jacobian, $\det(Q)^{-1}$. Thus when $f$ preserves volume so will the map in the new coordinates.}
\bsplit{transformVars}
	\xi &= P^T x ,\\
	\eta &= Q^{-1}\delta = Q^TD\Omega_*\delta . 
\esplit
In the new coordinates the map \Eq{ExpandGeneral} becomes
\bsplit{ExpandGeneralII}
	\xi' &
	      = \xi + \hat{\omega} + \kappa  R^T\eta' + \cO(\kappa^2) \mod 1 , \\
	\eta' &
	      = \eta + \kappa G(\xi) + \cO(\kappa^2) ,
\esplit
where, by \Eq{newFreq}, $\hat{\omega} = P^T \omega$ has the form \Eq{OmegaHat}, we used \Eq{PQR} to obtain
\[
	P^T D\Omega_* \delta = R^TQ^TD\Omega_*Q \eta = R^T \eta ,
\]
and we defined the transformed force
\beq{Fourier}
	G(\xi) \equiv Q^{-1} Y(P^{-T}\xi,y_*;0) = Q^{-1} \sum_{j\in\bZ^d} \tilde{Y}_{Pj} e^{2\pi i \xi^T j} ,
\eeq
which therefore has Fourier coefficients $\tilde{G}_j= Q^{-1}\tilde{Y}_{Pj}$. 


The map \Eq{ExpandGeneralII} has a natural slow-fast decomposition in terms of the (slow) resonant angles  $\xi_S= (\xi_1,\ldots \xi_r)^T$ and the (fast) nonresonant angles $\xi_F = (\xi_{r+1}, \ldots \xi_{d})^T$. Similarly we write the upper triangular matrix $R$ \Eq{PQR} in block form as $R = [ R_S \, R_F]$. 
Then since $s_r$ is an integer multiple of each $s_1,s_2,\ldots, s_r$, iterating \Eq{ExpandGeneralII} $s = s_r$ times to $\cO(\kappa^2)$ will eliminate the constant rational frequencies $\hat{n}_j/s_j$ in \Eq{OmegaHat}. Moreover, since the angles are taken mod $1$ we have
\begin{align*}
	\xi_S^s &= \xi_S^0 + \kappa s R_S^T \eta^s + \cO(\kappa^2) \mod 1 , \\
	\xi_F^s &= \xi_F^0 + s \hat\omega_F + \kappa s R_F^T \eta^s + \cO(\kappa^2) \mod 1 , \\
	\eta^s &= \eta^0 + \kappa \sum_{j=0}^{s-1} G(\xi + j\hat{\omega}) + \cO(\kappa^2) .
\end{align*}
If the vector $\hat\omega_F$ is Diophantine,  the averaging results of \cite[see \S5]{Dullin12a} imply that the fast angles $\xi_F$ can be averaged away on the time scale $\cO(\kappa^{-1})\gg 1$, 
leaving
\bsplit{AveragedSystem}
	\xi_S^s &= \xi_S^0 + \kappa s R_S^T \eta^s + \cO(\kappa^2) \mod 1 ,\\
	\eta^s &= \eta^0 + \kappa s \bar{G}(\xi_S^0) + \cO(\kappa^2) ,
\esplit
where the resonant force is
\bsplit{gbar}
	 \bar{G}(\xi_S) 
	&= \frac{1}{s} \int_{\bT^{d-r}} \sum_{j=0}^{s-1} G(\xi + j\hat{\omega})\,d\xi_F 
	= \frac{1}{s}\sum_{j=0}^{s-1} \sum_{l\in\bZ^d} \tilde{g_l} e^{2\pi ij \hat{\omega}\cdot l} 
	       \int_{\bT^{d-r}} e^{2\pi i \xi\cdot l} \,d\xi_F \\
	&= \sum_{l_S\in\bZ^r} \tilde{G}_{(l_S,0)} e^{2\pi i \xi_S\cdot l_S}
	           \frac{1}{s} \sum_{j=0}^{s-1}e^{2\pi ij \hat{\omega}_S\cdot l_S}
	=   \sum_{l_S \in \cL(\hat{\omega}_S)}
	       \tilde{G}_{(l_S,0)}e^{2\pi i\xi_S\cdot l_S} ,
\esplit
and
$
	\cL(\hat{\omega}_S) = \{l_S\in\bZ^r :\,\hat{\omega}_S\cdot l_S \in \bZ \}
$
is the $r$-dimensional resonance lattice for the slow, resonant frequencies. In \cite{Dullin12a} it was shown that this averaging transformation can repeated to $n^{th}$ order in such a way that the orbit of the averaged system for the slow variables is within $\cO(\kappa^{-n})$ of the slow projection of the true system for times up to $\cO(\kappa^{-1})$. 
Moreover, if the map is volume-preserving (symplectic), one can choose the transformation to also be volume-preserving (symplectic).

Since $R$ is upper triangular, the averaged map \Eq{AveragedSystem} has the form of a semi-direct product: the dynamics of $(\xi_S,\eta_S)$ do not depend on the nonresonant action variables, $\eta_F$, up to $\cO(\kappa^2)$.

If the map \Eq{ActionDepMap} were canonically symplectic in the original variables $(x,y)$, these actions (locally aligned with the resonance) are fixed, up to $\cO(\kappa^2)$: indeed whenever the force $Y(x,y_*,0)$ is a gradient its Fourier coefficients satisfy $\tilde{Y}_l \propto l$ so that, from \Eq{Fourier},
\[
	\tilde{G}_{(l_S,0)} \propto Q^{-1}P\left[\begin{array}{c} l_S \\ 0 \end{array}\right] = R\left[\begin{array}{c} l_S \\ 0 \end{array}\right] = R_Sl_S
\]
is always zero in the bottom $d-r$ components (again since $R$ is upper triangular). Of course, for this case, when $D\Omega_*$ is symmetric and positive definite, \Th{Nekhoroshev} implies much more: that the drift in all the actions is small for exponentially long times. On the other hand, if the force is not a gradient, then the nonresonant actions in \Eq{AveragedSystem} can experience forces that depend upon the slow angles $\xi_S$ and have a nonzero average, leading to large drifts.

\section{ACTION DRIFT IN RANK-ONE RESONANCES}\label{sec:RankOneDrift}

For a rank-one resonance, the averaged force \Eq{gbar} depends upon a single angle, $\xi_1 \propto m\cdot x$, and
the averaged system \Eq{AveragedSystem} reduces to
\bsplit{RankOneAverage}
	\xi_1^s &= \xi_1^0 + \kappa s \|p^{(1)}\|_{D\Omega_*}\,\eta_1^s + \cO(\kappa^2) \mod 1 , \\
	\eta^s &= \eta^0 + \kappa s \,\bar{G}(\xi_1^0) + \cO(\kappa^2),
\esplit
since, when $r=1$, we know
\[
	s R_{11} = s \| p^{(1)}\|_{D\Omega_*} = \|m\|_{D\Omega_*}
\]
where $p^{(1)}$ is the first column of $P$ in \Eq{SmithForm}. 

The first two components of \Eq{RankOneAverage} are a generalized Chirikov area-preserving map. These components reduce to the standard map when $\bar{G}$ has only a single Fourier component $l = \pm s$:
\[
	\bar{G}(\xi_1) = \tilde{G}_{(s,0)}e^{2\pi is\xi_1} + \tilde{G}_{(-s,0)}e^{-2\pi is\xi_1} = Q^TD\Omega_*( \tilde{Y}_{m}e^{2\pi i s\xi_1} +\tilde{Y}_{-m}e^{-2\pi i s\xi_1})
\]
so that $\bar{G}_1(\xi_1) =a_1 \cos(2\pi s\xi_1 + \theta)$ where 
\[
	a_1 = \frac{2}{\|m\|_{D\Omega_*}}\left| m^TD\Omega_* \tilde{Y}_m\right|\qquad\text{and}\qquad\theta = \arg(m^TD\Omega_* \tilde{Y}_m),
\]
since $q^{(1)} = m/\|m\|_{D\Omega_*}$.
When $a_1$ is small the dynamics of this map are dominated by the resonance island near $\eta_1 = 0$, recall \Fig{stdmapfli}. The half-width of this island is
\[
	\Delta\eta_1 = \sqrt{\frac2{\pi}\frac{sa_1}{\|m\|_{D\Omega_*}}}, 
\] 
but
$
	\Delta \eta_1 = \frac1{\kappa \|m\|_{D\Omega_*}}m^TD\Omega_*\Delta y,
$
so the half-width with respect to $y$ in the $D\Omega_* m$ direction (locally perpendicular to the resonance) is
\[
	w_m = 
	     \frac2{\|D\Omega_* m\|}\sqrt{\frac{s\eps}\pi\left|m^TD\Omega_* \tilde{Y}_m\right|} .
\]

Note that the the nonresonant actions in \Eq{RankOneAverage} are forced by the resonant angle $\xi_1$ through $\bar{G(\xi_2)}$. The components of this force that are out-of-phase with $\bar{G_1(\xi_1)}$ are responsible for the drift.
To see this, consider the special case that the first component of $\bar{G}$ is a pure sine function (e.g., $\theta \to \pi/2$), and that $s = \|p^{(1)}\| = 1$. Then, upon setting the formal parameter $\kappa \to 1$, the averaged dynamics of \Eq{RankOneAverage} become the $(1+d)$-dimensional slow map
\bsplit{rank1drift}
	\xi_1' &= \xi_1 + \eta_1' \mod 1 ,\\
	\eta_1' &= \eta_1 - \frac{a_1}{2\pi}\sin(2\pi\xi_1) ,\\
	\eta_j' &= \eta_j - \frac{1}{2\pi}\left(a_j\sin(2\pi \xi_1) + b_j\cos(2\pi \xi_1)\right),\quad j=2,\ldots,d .
\esplit
The standard map dynamics in $(\xi_1,\eta_1)$ then provides a oscillatory force on the remaining (drifting) actions, $\eta_2,\cdots,\eta_d$.
If $a_1\neq0$, then after $T$ iterations
\[
	\eta_j^{(T)} - \eta_j^{(0)} = -\frac{1}{2\pi}\sum_{t=0}^{T-1}a_j\sin(2\pi \xi_1^{(t)}) 
									+ b_j \cos(2\pi \xi_1^{(t)}) 
	           = \frac{a_j}{a_1}(\eta_1^{(T)}- \eta_1^{0}) 
	               - \frac{T}{2\pi} b_j \langle \cos(2\pi \xi_1)\rangle_T
\]
where
\beq{cosavg}
	\langle \cos(2\pi \xi_1)\rangle_T = \frac1{T}\sum_{t=0}^{T-1} \cos(2\pi\xi_1^{(t)})
\eeq
is the time average along the standard map orbit.
When $|a_1| \lesssim 0.971$, the critical parameter for the destruction of the last rotational invariant circle of the standard map, the resonant action $\eta_1$ is bounded. Thus, when $T\gg 1$ we have
\beq{AveDrift}
	\frac{\eta^{(T)}-\eta^{(0)}}{T} \approx -\frac{b}{2\pi} \langle \cos(2\pi\xi_1)\rangle_T
\eeq
Note that the long term drift is parallel to $b$, the vector of amplitudes of the out-of-phase forcing of the drift actions. For $T\gg 1$, $\langle \cos(2\pi \xi_1)\rangle_T$ will limit to a constant value determined by the decoupled standard map orbit. An example is shown in \Fig{StdMapAvg} for the case $a_1 = 0.52$. Note that inside the fixed point resonance there is a domain for which $|\xi_1^{(t)}|$ is small for all $t$, and thus $\cos(2\pi \xi_1^{(t)} ) > 0$, resulting in a positive time average. By contrast orbits near the separatrix of the resonance spend a majority of their time near the saddle fixed point at $(0.5,0)$, giving a negative average. Between these extremes there is a tube where $\langle \cos(2\pi\xi_1)\rangle_T\approx 0$. We will refer to this as \textit{bidirectional drift} in contrast to the unidirectional drift that would be caused by a constant forcing term. We saw bidirectional drift in \Fig{OrbitProjections}(b)---the three orbits correspond to the variation in direction as one changes diameter of the resonant dynamics.


\InsertFig{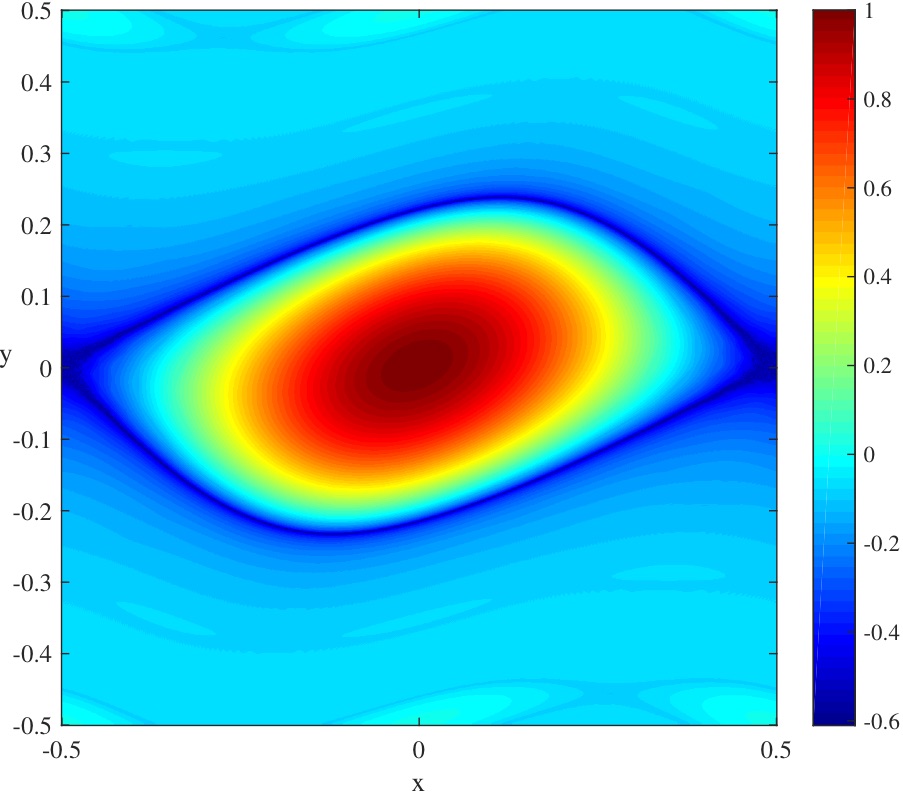}{Time average of $\cos(2\pi \xi_1)$, \Eq{cosavg}, for orbits of the standard map for $a_1=0.52$, with $10^6$ initial conditions on a square grid, each iterated $10^3$ steps}{StdMapAvg}{4in}

Thus in addition to the (symplectic) case where $a_j = b_j = 0$ for $j\geq 2$, for small nonzero $a_1$, in order to prevent drift, to lowest order, it is sufficient that $b = 0$, or that all drift action forcing is in phase with the resonant action forcing. One can also study higher order terms in the expansion about a resonance to understand the change in direction of the drift as the action changes \cite{Guillery18}.


\section{DRIFT IN THE GENERALIZED FROESCHL\'E MAP}\label{sec:FroDrift}

As the simplest example, we return to the generalized \fro map \Eq{GenFroMap} on $\bT^2 \times \bR^2$, with force and frequency map \Eq{GFForce}. The amplitudes $a$, $b$, and $c$ create the $(1,0,n)$, $(0,1,n)$ and $(1,1,n)$ resonances respectively.

Suppose first that $y_*$ is a point in an $(1,1,n)$ resonance, so that $y_{1*} + y_{2*} = n$.
Since $D\Omega = I$, the transformation matrices in \Eq{PQR} become
\[
	P = \begin{pmatrix}  1 & 1 \\ 1 & 2 \end{pmatrix}
	  =  QR = \frac{1}{\sqrt{2}} \begin{pmatrix} 1 & -1 \\ 1 & 1 \end{pmatrix}
	  			\begin{pmatrix}  \sqrt{2} & 3/\sqrt{2} \\ 0 & 1/\sqrt{2} \end{pmatrix} ,
\]
so that the new variables of \Eq{transformVars} are
\begin{align*}
	(\xi_1, \xi_2) &= (x_1+x_2, x_1 + 2x_2) , \\
	(\eta_1,\eta_2) &= \tfrac{1}{\sqrt{2}\kappa} ( y_1+y_2-n, y_2-y_1 + y_{1*}-y_{2*}) ,
\end{align*} 
and the new frequency, \Eq{newFreq}, is $\hat{\omega} = (\omega_S,\omega_F) = (n, y_{1*} + 2y_{2*})$. By periodicity and the resonance condition, the constant frequency for angle $\xi_1$ vanishes mod 1, so the transformed map, \Eq{ExpandGeneralII},  becomes
\begin{align*}
	\xi_1' &= \xi_1 + \kappa \sqrt2\eta_1' \mod 1\\
	\xi_2' &= \xi_2 + \omega_F + \cO(\kappa ) \mod 1\\
	\eta' &= \eta - \kappa\frac{c}{\sqrt{2}\pi}\left[\begin{array}{c} \cos(\pi\vphi)\sin(2\pi(\xi_1 + \vphi/2)) \\ \sin(\pi\vphi)\cos(2\pi(\xi_1 + \vphi/2)) \end{array}\right] + \cO(\kappa a,\kappa b)
\end{align*}
If the second frequency $\omega_F$ is Diophantine, then for $\kappa$ small enough we can apply the averaging procedures to push the $\xi_2$ dependence in the $a,b$ forcing terms to $\cO(\kappa ^2)$. Defining a final shifted angle $\xi_1 = \xi_1 + \vphi/2$, at first order we then obtain a slow map of the form \Eq{rank1drift} with 
\[
	a_1\propto c\cos(\pi\vphi),\qquad\qquad a_2 = 0,\qquad\qquad b_2 \propto c\sin(\pi\vphi)
\]
which has orbits that are conjugate to $\cO(\kappa^2)$ to the projection of an orbit of the original map onto the slow variables for times $\cO(\kappa^{-1})$. The drift in $\eta_2$ vanishes for the symplectic case $\vphi = 0$, but otherwise $b_2\neq0$. The result is bidirectional drift in $(1,1)$ resonances, as discussed in \Sec{RankOneDrift}. 

A comparison of the theory with computations is given in \Fig{GenFroeschleDriftCo11}. To do this, we compute the exact dynamics for an orbit that starts near the elliptic point of the $(\xi_1,\eta_1)$ dynamics in the $(1,1,0)$ resonance. Thus we set $(x,y) = (-\tfrac12\vphi,0,-0.35,0.35)$ when $0 \le\vphi <\tfrac12$ and $(x,y) = (\tfrac12(1-\vphi),0,-0.35,0.35)$ when $\tfrac12 <\vphi \le 1$ since the sign of $a_1$ changes at $\vphi = \tfrac12$, where the amplitude of the resonance vanishes to $\cO(\kappa)$. After iterating $10^3$ steps we compute a linear fit to the values of $\eta_2^{(t)}$ to obtain the slope; this gives the drift shown in the figure as a function of $\vphi$. The numerical points agree very well with the theoretical drift from \Eq{AveDrift}, also shown---except for a small deviation near the discontinuity at $\vphi = \tfrac12$.

\InsertFig{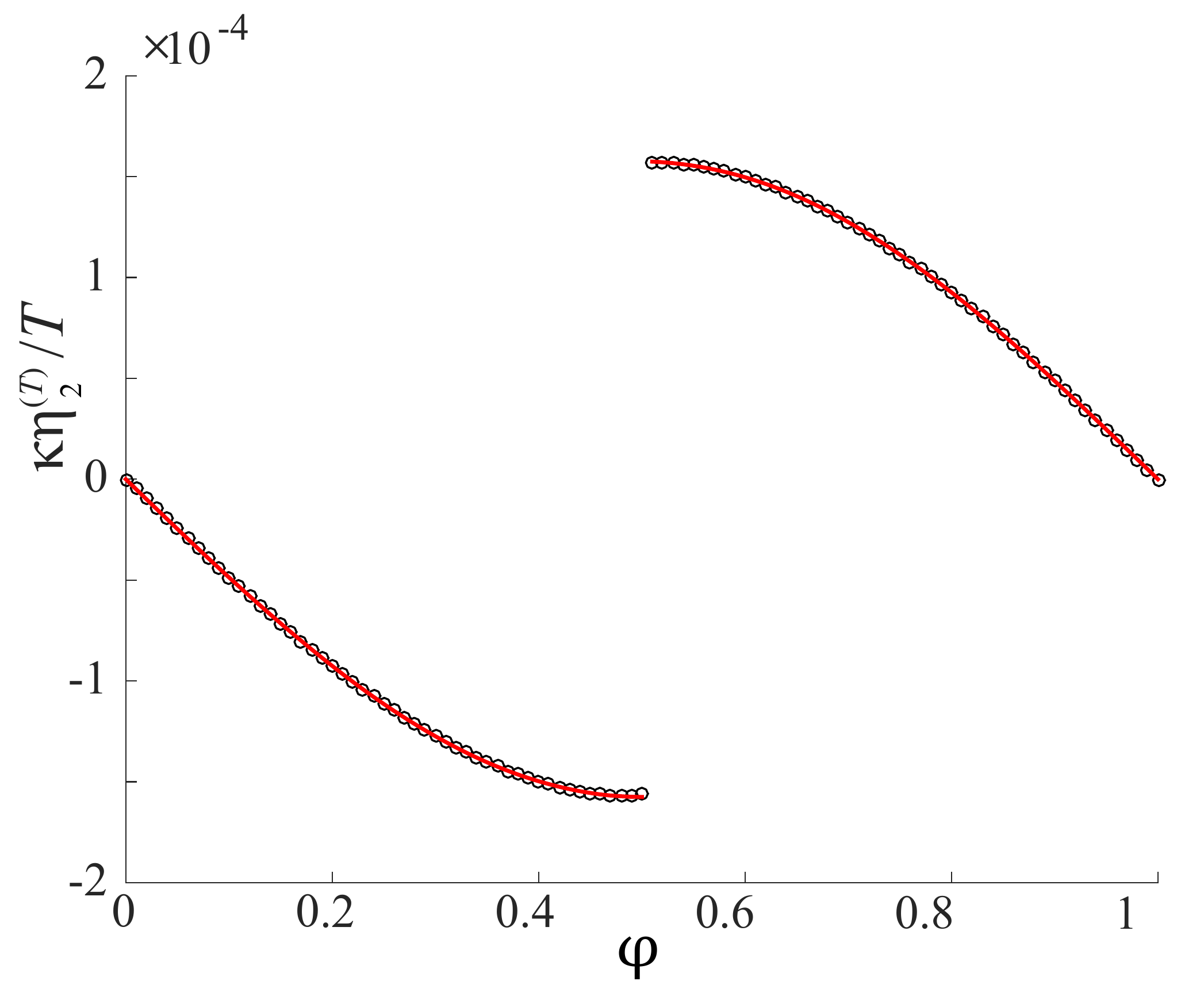}{Comparison of the computed drift in $\kappa \eta_2^{(t)}$ to the analytical drift from \Eq{AveDrift} in the $(1,1,0)$ resonance with actions $y^{(0)} = (-0.35,0.35)$ iterated $T = 10^3$ steps, for the generalized \fro map with $\eps(a,b,c) = (0.001,0.001,0.0007)$. The curve (red)
shows the analytical drift $-\kappa \frac{c}{\sqrt{2}\pi}\sin(\pi \vphi) \sgn(\cos(\pi\vphi))$ from \Eq{AveDrift}.}{GenFroeschleDriftCo11}{4in}

Consider now a point $y_*$ in an $(1,0,n)$ resonance, $m\cdot \Omega(y_*) = y_{1*} = n$ where $\Omega$ is the identity map. Since $\xi_1 = m\cdot x = x_1$ the map \Eq{GenFroMap} is already written in resonant coordinates, once we define $\delta = (y-y_*)/\kappa$:
\begin{align*}
	x_1' &= x_1 + \kappa \delta_1' \mod 1 ,\\
	x_2' &= x_2 + \omega_F + \cO(\kappa) \mod 1 ,\\
	\delta' &= \delta -  \frac{\kappa}{2\pi}\left[\begin{array}{c} 
						a \sin(2\pi x_1) \\ 
	                    0 \end{array}\right] + \cO(\kappa b,\kappa c) .
\end{align*}
Here the  fast frequency is $\omega_F = y_{2*}$ and, since this is almost always irrational, the resonance has rank one. Under first order averaging, the fast $x_2$ dependence in the $b,c$ forcing terms is zero to $\cO(\kappa^2)$, so that $\delta_2$ is conserved to $\cO(\kappa^2)$ on the time scale $t\sim\cO(\kappa^{-1})$. However, we can carry out the averaging procedure to two more orders to remove the fast dependence up to $\cO(\kappa^4)$ and compute a nonzero drift. This procedure gives
\bsplit{thirdOrder}
	\xi_1' &= \xi_1 + \kappa \eta_1' \mod 1 ,\\
	\eta_1' &= \eta_1 - \frac1{2\pi}\left(\kappa a - \kappa^3 \frac{bc}{8\sin^2(\pi\omega_F)}\right)\sin(2\pi \xi_1) - \kappa^3 \frac{c^2}{16\pi\sin^2(\pi\omega_F)}\sin(2\pi\vphi) ,\\
	\eta_2' &= \eta_2+ \kappa^3 \frac{c^2}{16\pi\sin^2(\pi\omega_F)}\ \sin(2\pi\vphi) .
\esplit
Note that in this case the drift term in $\eta_2$ is independent of the slow variables. For $\vphi\neq0$ or $\tfrac12$, this implies unidirectional drift in $(1,0,n)$ resonances that is second order in $\eps$ with respect to the original variables. The drift speed depends on $\omega_F$ and is nonzero and finite except at a double resonance when $\omega_F \in \bZ$.  By symmetry, a similar second order drift exists for orbits in $(0,1,n)$ resonances as well.

We show a comparison of this second order drift with computations in \Fig{GenFroeschleDriftCo10}. Panel (a) shows the variation in drift with $\vphi$, and (b) with $\omega_F$. Again the analysis agrees with the numerical computations, though there is a visible deviation between the theory and computations as a function of the phase shift.

\InsertFigTwo{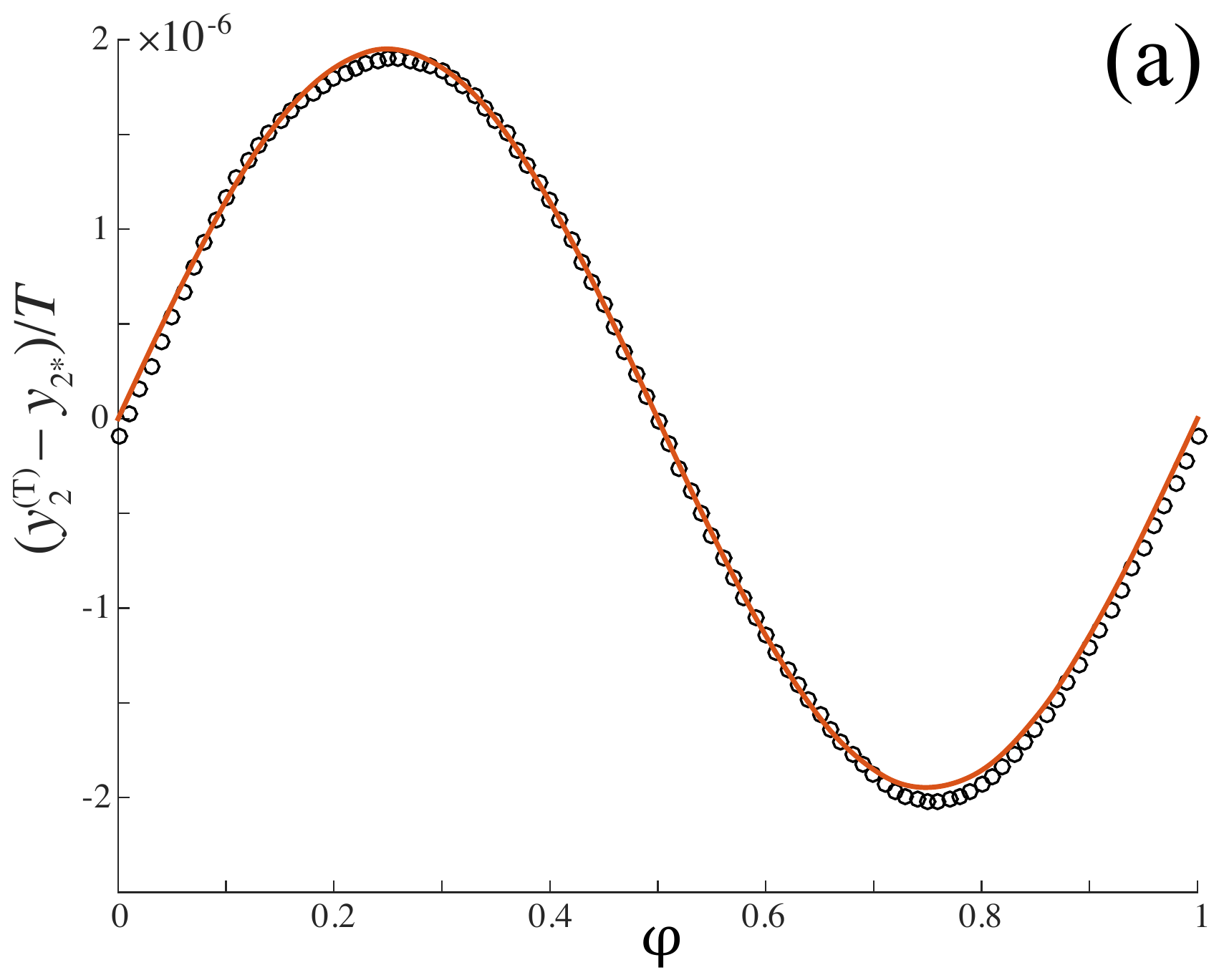}{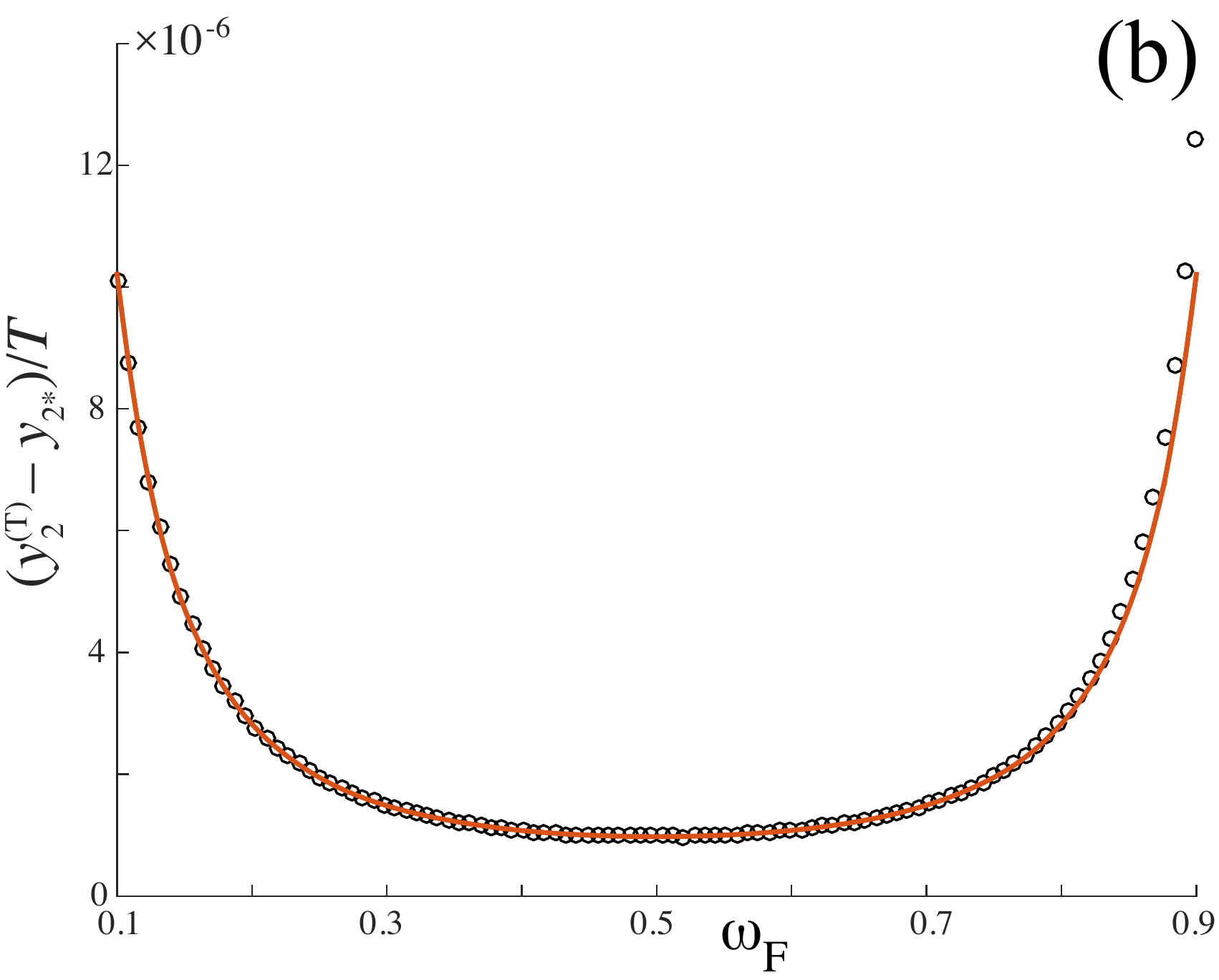}{Comparison of the drift in $y_2^{(t)}-y_{2*}$ to the analytical drift from the third order averaged map \Eq{thirdOrder} for an orbit of length $T= 10^3$, starting at $(x,y) = (0,0,0,y_{2*})$ in the $(1,0,0)$ resonance of the generalized \fro map with \Eq{GFForce}. Here $\eps(a,b,c) = (0.01,0.01,0.007)$. (a) Variation of the drift with $\vphi$ when $\omega_F = y_{2*} = 0.25$. (b) Variation of the drift as a function of $\omega_F = y_{2*}$ when $\vphi =0.25$. The curves show the analytical drift $\kappa^4 c^2 \sin(2\pi\vphi)/(16\pi \sin^2(\pi \omega_F))$.}{GenFroeschleDriftCo10}{2.5in}

\section{CONCLUSIONS}\label{sec:Conclusions}

Action-angle maps, like \Eq{ActionDepMap} are natural as descriptions of the dynamics of nearly integrable systems. When such a map is symplectic and has a positive-definite twist, it has a strong quasi-stability property due to Nekhoroshev's theorem: for $\eps \ll 1$ the actions are confined for exponentially long times. By contrast when the map is merely volume preserving, we showed the actions can drift along resonances much more rapidly than the exponentially slow diffusion of a symplectic map. The is due to the simple geometrical fact that when the symplectic condition is relaxed the resonant forces are no longer orthogonal to the resonance manifold $m \cdot \Omega(y)  = n$. The average dynamics show that the drift speed along the resonance has a sign that depends on whether the resonant subsystem is trapped in the elliptic island or is in the chaotic separatrix region of the resonance; the drift is ``bidirectional".

For the generalized \fro map \Eq{GenFroMap} that we studied in this paper, with the simple force $F(x)$ and frequency map $\Omega(y)$ \Eq{GFForce}, the nonsymplectic term is---to lowest order---only important in $(m,n) = (1,1,n)$ resonances since the nonsymplectic coupling term is slowly varying in these resonances. Of course, to higher order, as noted in \Sec{FroDrift}, nonlinearity of the map can also cause drift in other resonances.

\InsertFig{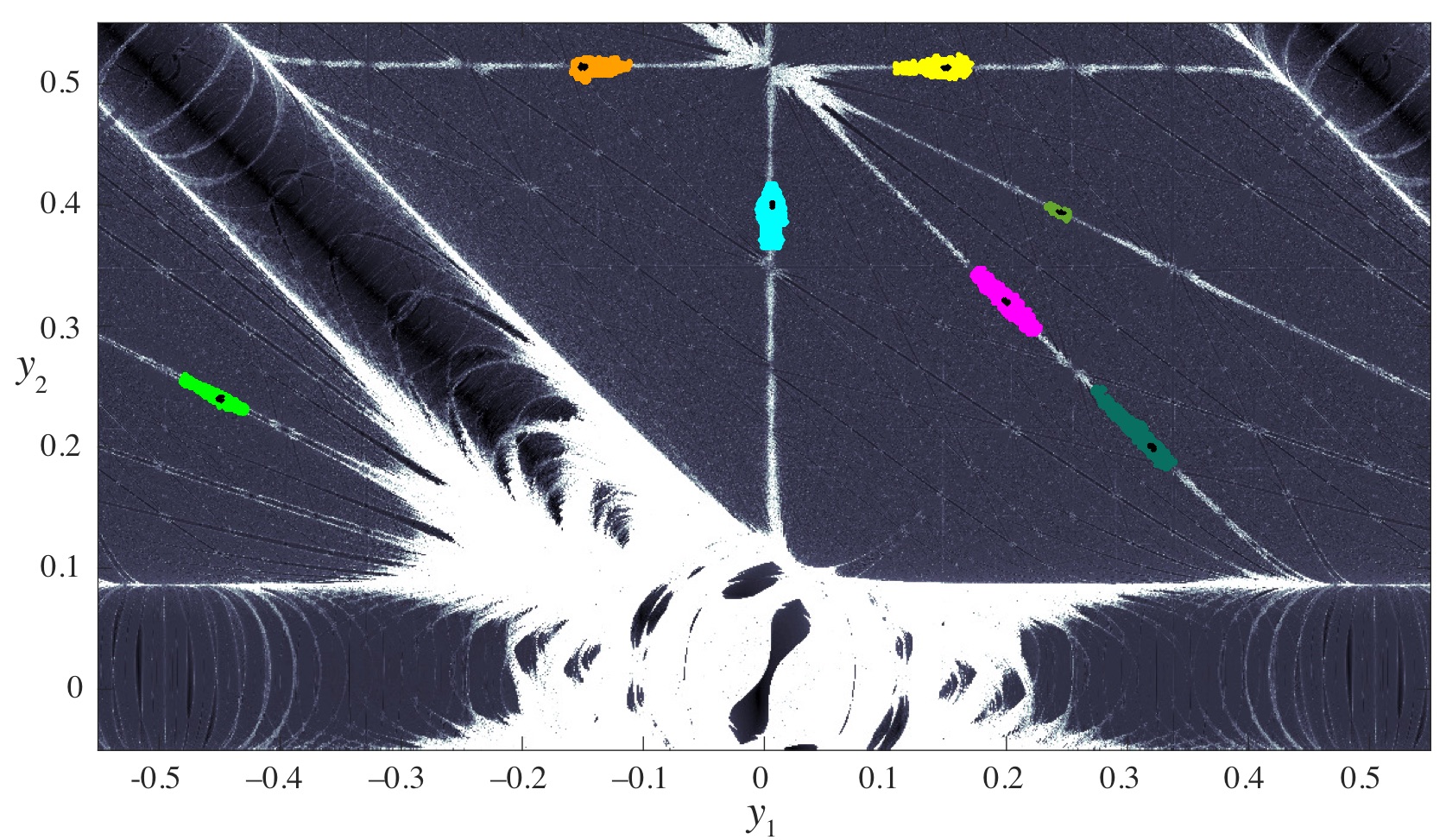}{Drift in the symplectic generalized \fro map with \Eq{GFForce} and parameters $(a,b,c) =  (0,0.1,0.07)$, $\vphi = 0.0$ as well as an added full spectrum force \Eq{FFullSpec} with $d = 10^{-4}$ with $\vphi_{fs} = 0.0$.
The grayscale shows the FLI for initial conditions in the plane $(0,0.25,y_1,y_2)$. Also shown the points on seven clusters of orbits of length $T=10^8$ that have the largest values of the FLI near the initial value (black dots). Each cluster is 100 orbits, and points are shown only if they are within $10^{-3}$ of the slice.}{FullSpectrumSym}{5in}

It is interesting, however, to consider the effect of more general forces. For example, the force
\beq{FFullSpec}
F_{fs} = -\frac{1}{2\pi}\frac{d}{(2.1+\cos(2\pi x_1) + \cos(2\pi x_2))^2} 
		\begin{pmatrix}
     		\sin(2\pi x_1)  \\
     		\sin(2\pi(x_2 + \vphi_{fs}))
		\end{pmatrix}
\eeq
has been studied in \cite{Froeschle06} for the symplectic case $\vphi_{fs} = 0$, where it is a gradient. This force has a full spectrum of Fourier coefficients.
When $\vphi = \vphi_{fs} = 0$, the drift of the actions in a map with  force $F(x) + F_{fs}(x)$ is still constrained by Nekhoroshev's theorem since the map is symplectic and the twist is positive definite. Figure~\ref{fig:FullSpectrumSym} shows the FLI (in grayscale as in \Fig{OrbitProjections}) on the plane $x = (0,0.25)$. Note that there are many more (small) resonance channels visible in this FLI plot than there were for the force \Eq{GFForce} alone. Also shown in the figure are points on seven clusters of orbits, each started near the small black dots in the figure. Each of these clusters consist of the $100$ orbits that have the largest FLI values in the neighborhood of the initial point. The orbits are iterated $10^8$ times, but points are plotted only when they fall within $10^{-3}$ of the plane. Note that the drift in actions for each of these clusters is small: the total change is less than $\cO(0.1)$.

By contrast, when $\vphi$ or $\vphi_{fs} \neq 0$, the drift can be large, even in narrow resonance channels. An example is shown in \Fig{FullSpectrumVP} for a map with the same parameters as that in \Fig{FullSpectrumSym} except that $\vphi_{fs} = 0.5$. The FLI for this volume-preserving system is hard to distinguish from the symplectic case. However, the points on the single cluster of orbits shown now exhibit drifts with size $\cO(1)$ in $10^8$ steps. Note also that the drifting orbit moves from one rank-one resonance channel to another when it reaches a rank-two crossing.  We plan to investigate these transitions in a future paper. 

\InsertFig{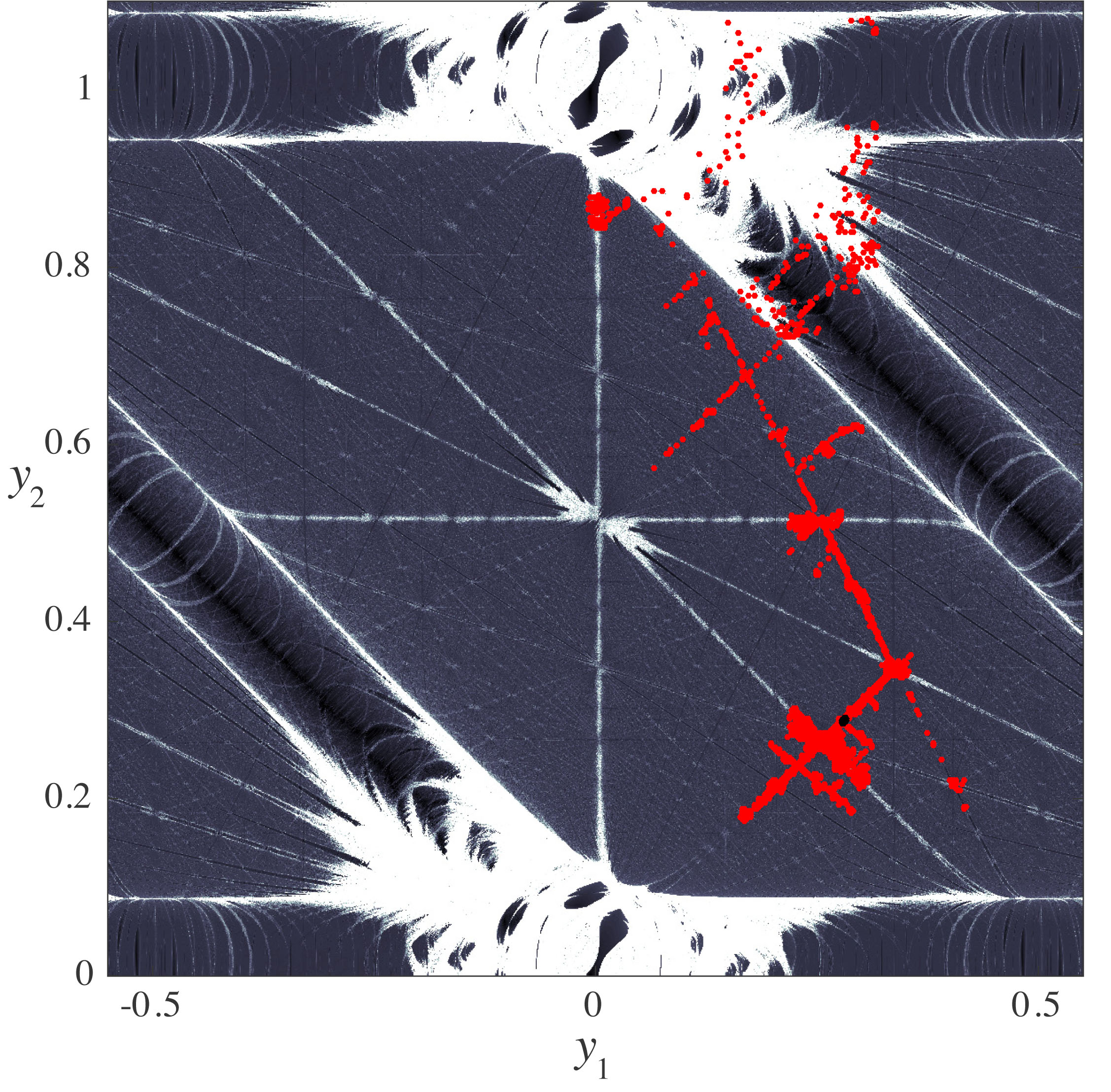}{Drift in a volume-preserving, full-spectrum, generalized \fro map with the same parameters as \Fig{FullSpectrumSym} except that $\vphi_{fs} = 0.5$. The grayscale shows the FLI for initial conditions in the plane $(0,0.5,y_1,y_2)$. Also shown the points (red online) on a cluster of $100$ orbits of length $T=10^8$ that have the largest FLI values near the initial point $(x,y) = (0,0.5,0.28,0.288)$ (black dot). }{FullSpectrumVP}{4.in}

\appendix

\section*{APPENDIX A. REVERSIBILITY} \label{app:Reversible}

An invertible map $f$ is \emph{reversible} if there exists an involution $S$ ($S^2=I$) such that 
\beq{Reversible}
	S\circ f = f^{-1} \circ S.
\eeq
The map $S$ is called a \emph{reversing symmetry} for $f$ \cite{Lamb98}. 
The generalized \fro map \Eq{GenFroMap} for any $d$ and $k$ is reversible whenever the force is odd, $F(-x) = -F(x)$, under the reversor
\[
	S_1:\left\{\begin{array}{ccl}
						x' & = & -x,\\
						y' & = & y+\eps F(x).
					\end{array}\right.
\]
Similarly, if the frequency map is odd, $\Omega(-y) = -\Omega(y)$, $f$ has the reversor
\[
		S_2:\left\{\begin{array}{ccl}
						x' & = & x,\\
						y' & = & -y-\eps F(x).
					\end{array}\right.
\]
These reversors can be generalized to functions that are odd about another point, besides the origin \cite{Fox14}.

The composition of a symmetry and a reversor is a reversor, and the collection of symmetries and reversors forms the ``reversing symmetry group." The map $f$ has two obvious symmetries. First $f$ is a symmetry of itself; so, if $S$ is a reversor, so is $f^t \circ S$, for any $t \in \bZ$. The second symmetry is induced by the rotation $R_m(x,y) = (x+m,y)$ for any $m \in \bZ$. If we lift $f$ to $\bR^{d+k}$, then $f\circ R_m = R_m \circ f$. This implies that $S \circ R_m$ is a reversor.

A periodic orbit is ``symmetric" if it is invariant under a reversor $S$. In such a case the orbit necessarily has points on either fixed set $\Fix{S} = \{(x,y): S(x,y) = (x,y)\}$ or
$\Fix{f \circ S}$. \cite{Kook89}. For example, the fixed sets of for the $S_1$ reversor are $d$-manifolds $\text{Fix}(S_1) = \{x=0\}$ and $\text{Fix}(f\circ S_1) = \{\Omega(y)=2x\}$. The gives additional reversors with the fixed sets $\text{Fix}(S_1\circ R_m) = \{x=\tfrac12 m\}$ and $\text{Fix}(f\circ S_1\circ R_m) = \{\Omega(y)=2x-m\}$ that are helpful in classifying the periodic orbits.

\section*{APPENDIX B. STABILITY AND REFLEXIVITY}\label{app:Stability}

Recall that linear stability is determined by the eigenvalues of the product of the Jacobians $Df(x^{(t)},y^{(t)})$ evaluated along the orbit. For the generalized \fro map \Eq{GenFroMap}, the Jacobian is \Eq{Df}.
We call these eigenvalues the ``multipliers" of the orbit.

An important property of symplectic maps with a constant Poisson matrix (like $J$ in \Eq{Poisson}) is that the multipliers of any periodic orbit have the reflexive property: whenever $\lambda$ is an eigenvalue, then $\lambda^{-1}$ is also an eigenvalue with the same multiplicity. This follows from the the symplectic condition \Eq{Poisson}, which, upon differentiation gives
\[
	J Df J^{-1} = Df^{-T} .
\]
Thus $Df$ is conjugate to its inverse transpose, implying that $Df$ and $Df^{-1}$ have the same characteristic polynomials.

Reflexivity holds also for symmetric orbits of reversible maps. Differentiation of the conjugacy \Eq{Reversible} between $f$ and $f^{-1}$ gives
\[
	DS(f(z)) Df(z) = Df^{-1}(S(z)) DS(z)
\]
so that if $z = (x_*,y_*)$, is a fixed point of both $f$ and $S$, then $Df$ is conjugate to $Df^{-1}$.
Note that the linearization of the generalized \fro map \Eq{GenFroMap} about a fixed point is reversible for \textit{any} dimensions $d$ and $k$ and \textit{any} force and frequency map (even if the map itself is not globally reversible). Indeed, the linearization is the map
\begin{align*}
		\delta x' &= x + D\Omega(y_*) \delta y' ,\\
		\delta y' &= y + \eps DF(x_*) \delta x ,
\end{align*}
which has the same form as \Eq{GenFroMap} with the odd force $DF \delta x \to F(\delta x)$ and odd frequency map $D\Omega \delta y \to \Omega(\delta y)$. Thus, by the results of App. A, this map is reversible with, e.g., the linear reversor $S_1(\delta x, \delta y) = (-\delta x , \delta y + \eps DF \delta x)$. Consequently the eigenvalues of the matrix \Eq{Df} have the reflexive property at any point $(x,y)$.

For example for $d=k=2$, the characteristic polynomial for \Eq{Df} at a fixed point is
\bsplit{CP}
	p(\lambda) &= \det(Df-\lambda I) = \lambda^4 - A \lambda^3 + B\lambda^2 - A\lambda +1, \\
	A          &=  \Tr(Df) =  4 + \Tr(D\Omega DF)\\ 
	B          &=  \tfrac12 [ \Tr(Df)^2 - \Tr(Df^2)]
	            = 6 + 2\Tr(D\Omega DF) + \det(DF) \det(D\Omega) .
\esplit
Since the coefficients of $\lambda^3$ and $\lambda$ are the same in the characteristic polynomial, whenever $p(\lambda) = 0$ then $p(\lambda^{-1})=0$ too.
Therefore we cannot distinguish the properties of the generalized map \Eq{GenFroMap} from the symplectic case on the basis of eigenvalues of its fixed points alone.

However, it is easy to see by example that when $DF$ or $D\Omega$ is not symmetric, then the reflexive property does not generally hold for orbits with period larger than one. The implication is that the map \Eq{GenFroMap} cannot be symplectic with a constant Poisson matrix.

\section*{APPENDIX C. SMOOTHING}\label{app:Smoothing}

To compute the drift of action, we use exponential smoothing. Given a sequence $\{z^{(t)}: t =0,1,\ldots\}$, we define 
\beq{singleSmooth}
	\smooth{z} \equiv N \sum_{s=0}^{T} z^{(T-s)} e^{-\lambda s}, \quad N = 1-e^{-\lambda} .
\eeq
Here the normalization factor $N$ is appropriate when $T \gg \lambda^{-1}$, since
\[
	\smooth{1} = 1-e^{-\lambda (T+1)} \quad \xrightarrow[T \to \infty]{} \quad 1.
\]
This process gives the majority of the weight for the average to the most recent data; 
for instance if $\lambda = 0.01$, then the most recent 100, 200, and 300 iterates are given $63\%,86\%,$ and $95\%$ of the weight respectively.
The smoothing \Eq{singleSmooth} can be easily computed iteratively by
\beq{iterativeSmooth}
	\smooth[t]{z} = N z^{(t)} - (1-N) \smooth[t-1]{z}, \quad 
		\smooth[-1]{z} \equiv 0.
\eeq

For example, applying this to the changes in angle $z^{(t)}= x^{(t+1)}-x^{(t)}$ gives a running average of the rotation vector $\omega^T$.
When $\Omega$ is the identity map in \Eq{GenFroMap} $x^{(t+1)}- x^{(t)} = y^{(t)}$, so this is equivalent to a smoothed action $ J^T =\smooth{y}$.

For maps with drifting orbits, we want to obtain a linear trend for the actions. This can be done with
the ``Holt-Winters" double exponential smoothing \cite{Gardner06} which computes two smoothed variables, a mean $J^{(t)}$ and a trend $V^{(t)}$, using the iteration
\bsplit{HoltWinters}
	J^{(t)} 	&= N_1 y^{(t)} + (1 - N_1) (J^{(t-1)} + V^{(t-1)}) \\
	V^{(t)} 		&= N_2 (J^{(t)} - J^{(t-1)}) + (1 - N_2) V^{(t-1)}
\esplit
where $N_i = 1-e^{-\lambda_i}$ for exponents $\lambda_1$ and $\lambda_2$. Here one typically takes $J^{(1)} = y^{(1)}$, and $V^{(1)} = y^{(1)} - y^{(0)}$,
The forecasted trend at time $t$ is given by $\langle{y}\rangle^{t+s} = J^{(t)} + s V^{(t)}$.

For our computations we typically set $\lambda_1 = \lambda_2 = 0.01$.

\section*{ACKNOWLEDGMENTS}
 The authors were supported in part by NSF grant DMS-1211350. We thank A. B\"acker, R. Easton,  and R. Ketzmerick for helpful conversations, and J. Laskar for posing an interesting question.

\bibliographystyle{alpha}
\bibliography{VPDrift}

\end{document}